\documentclass[3p,times,twocolumn]{elsarticle}
\usepackage{ecrc}

\volume{00}
\firstpage{1}
\journalname{Procedia Computer Science}
\runauth{}
\jid{procs}
\jnltitlelogo{Procedia Computer Science}

\usepackage{amssymb}
\usepackage[figuresright]{rotating}
\usepackage{natbib}
\usepackage{tikz}
\usepackage{amsmath,amssymb}
\usepackage[toc,page]{appendix}
\usepackage{pgfplotstable}
\pgfplotsset{compat=1.17}
\usepackage{booktabs}
\usepackage{subcaption}
\usepackage{xcolor}
\usepackage{enumitem} 
\usepackage{url}
\usepackage[justification=centering]{caption}
\usepackage{hyperref}
\usepackage{bbm}
\usepackage{setspace}
\usepackage{etoolbox}
\usepackage{graphicx} 

\AtBeginEnvironment{quote}{\par\singlespacing\small}

\newcommand{\splitatcommas}[1]{%
  \begingroup
  \begingroup\lccode`~=`, \lowercase{\endgroup
    \edef~{\mathchar\the\mathcode`, \penalty0 \noexpand\hspace{0pt plus 1em}}%
  }\mathcode`,="8000 #1%
  \endgroup
}

\begin{document}

\begin{frontmatter}

    \title{
        M6 Investment Challenge:\\
        The Role of Luck and Strategic Considerations
    }

    \author{
        Filip Staněk\fnref{cor1}
    }
    \address{CERGE-EI}
    \fntext[cor1]{
        \texttt{filip.stanek@cerge-ei.cz}\\
        CERGE-EI, a joint workplace of Charles University and the Economics Institute of the Czech Academy of Sciences, Politickych veznu 7, 111 21 Prague, Czech Republic.
    }

    \dochead{}

    \begin{abstract}

        This article investigates the influence of luck and strategic considerations on performance of teams participating in the M6 investment challenge.
        We find that there is insufficient evidence to suggest that the extreme Sharpe ratios observed are beyond what one would expect by chance, given the number of teams, and thus not necessarily indicative of the possibility of consistently attaining abnormal returns. 
        Furthermore, we introduce a stylized model of the competition to derive and analyze a portfolio strategy optimized for attaining the top rank.
        The results demonstrate that the task of achieving the top rank is not necessarily identical to that of attaining the best investment returns in expectation.
        It is possible to improve one's chances of winning, even without the ability to attain abnormal returns, by choosing portfolio weights adversarially based on the current competition ranking.
        Empirical analysis of submitted portfolio weights aligns with this finding.

    \end{abstract}

    \begin{keyword}
        M6 Forecasting Competition, Investment Competition, Portfolio Optimization, Dynamic Programming\\
        \emph{JEL codes:} G11, G17, G41, C61

    \end{keyword}

\end{frontmatter}

\section{Introduction}

This article analyzes factors influencing the performance of teams participating in the investment challenge of the M6 forecasting competition \citep[see][]{makridakisM6ForecastingCompetition2023}.
By employing a combination of statistical analyses and model-based simulations, we assess the extent to which the final ranking can be attributed to luck and how teams might have benefited from recognizing the nonlinear reward structure and adversarial nature of the competition when constructing their portfolios.

The first part of our analysis examines the role that random variation played in the final ranking and the possibility of consistently achieving abnormal returns.
We develop a stylized model of the competition that replicates key aggregate statistics and use it to assess the properties of the \citet{wrightTestEqualityMultiple2014} test of equality of expected Sharpe ratios under conditions observed in the M6 competition.
After verifying its applicability, we employ the test to assess whether the variation in Sharpe ratios observed among teams necessarily implies differences in expected Sharpe ratios, or whether it can be explained merely by random variation.
Despite the large variation in Sharpe ratios, the results do not appear to be incompatible with the null hypothesis of equality of expected Sharpe ratios.
Additionally, we use the stylized model to assess the extent to which the ability to attain abnormal returns improves one’s chances of securing different ranks in the investment challenge.

The second part of our analysis focuses on the strategic aspects inherent to the competition format, which might motivate teams to pursue objectives beyond merely maximizing Sharpe ratios. 
Extensive literature documents the effects of competition-like incentive structures on investment decisions, especially regarding fund managers' compensation and behavior. 
\citet{brownTournamentsTemptationsAnalysis1996} show that when fund managers' compensation is linked to relative performance, those performing poorly mid-year tend to choose more volatile portfolios in the latter half of the evaluation period compared to mid-year leaders.
Similarly, \citet{eltonIncentiveFeesMutual2003} find that funds with performance-linked compensation increase risk following a period of poor performance.
\citet{krasnyAssetPricingStatus2011} shows that in contexts where investors are concerned with relative wealth, those performing poorly previously are likely to seek portfolio choices that maximize chances of improving their rank, whereas successful investors aim to maintain the status quo and hedge against the choices of others.
In experimental settings, \citet{linFundConvexityTail2011} and \citet{dijkRankMattersImpact2014} show that contestants adapt their portfolios to their current relative performance, selecting assets with either positively or negatively skewed returns to maximize the probability of securing or retaining a top rank, respectively.
Particularly relevant in the context of M6, \citet{niekenRisktakingTournamentsTheory2010} demonstrate that the behaviors of both top-ranking contestants and contenders critically depend on the correlation of the returns from available investment strategies.
A contender is motivated to choose a strategy dissimilar to that of the top-ranking contestant to maximize the probability of significantly outperforming him/her, whereas the top-ranking contestant is motivated to mimic the strategy of the contender to minimize the probability of being overtaken.

To assess to what degree such effects might be relevant to the M6 investment challenge, we formulate the problem of maximizing the probability of securing a top rank by manipulating the proportion of short positions and, as a result, the degree of correlation between one's returns and the returns of other teams, as a dynamic programming problem and numerically solve it in the stylized model of the competition introduced earlier.
We then let this optimal strategy repeatedly compete in the stylized model of the competition to assess to which degree teams might have benefited by recognizing the competitive nature of the challenge when constructing their portfolios.
Strategically adjusting the portfolio based on current rankings substantially improves the probability of achieving a top rank.
To confirm the robustness of these results, we also replicate the exercise in an environment where returns and portfolio weights of competitors are bootstrapped directly from those in the M6 investment challenge, confirming our initial results.
An examination of submitted portfolio weights corroborates the findings from our simulations and bootstrap exercises.

The remainder of this article is organized as follows.
Section \ref{subsection:problem_setup} introduces the competition's rules and notation.
Section \ref{section:role_of_luck} tests the equality of expected Sharpe ratios and assesses to what extent the observed results are attributable to random variation.
Section \ref{section:role_of_strategic} derives the optimal strategy for attaining a top rank, tests its performance in simulated and bootstrapped environments, and reviews empirical evidence from submitted portfolios to validate the simulation outcomes.
Section \ref{section:conclusions} concludes.\footnote{
    A replication repository for this article is available at \url{https://github.com/stanek-fi/M6_Luck_and_Strategic_Considerations}. 
}

\section{Problem Setup}\label{subsection:problem_setup}

The competition includes a universe of $I=100$ assets, consisting of 50 stocks from the S\&P 500 and 50 ETFs. 
There are $K\in \mathcal{N}$ teams (each team is composed of up to 5 individuals) competing in either quintile prediction for these assets (forecasting challenge), portfolio optimization (investment challenge), or a combination of both (duathlon challenge).

In the investment challenge, each team $k$ is tasked with submitting portfolio weights $w_{i,m,k}$ for assets $1 \leq i \leq I$ over the span of $M=12$ four-week long intervals $T_{m}$.
Portfolio weights are submitted prior to the beginning of each interval $T_{m}$ with $1 \leq m \leq M$, and the sum of the portfolio weights in absolute value is restricted to the interval:
\begin{equation}
    0.25 \leq \sum_{i=1}^{I}|w_{i,m,k}|\leq 1.
\end{equation}
The plain investment returns of team $k$ at day $t \in T_m$ in interval $m$, denoted as $ret_{t,k}$, are computed as:
\begin{equation}
    RET_{t,k}= \sum_{i=1}^{I} w_{i,m,k} r_{i,t} \quad \text{with} \quad r_{i,t}=\frac{S_{i,t}}{S_{i,t-1}}-1,
\end{equation}
\begin{equation}\label{eq:log_transform}
    ret_{t,k}= ln(1+RET_{t,k}),
\end{equation}
where $S_{i,t}$ denotes the asset price for asset $1 \leq i \leq I$ at day $t$.
The standardized investment returns from $t_1$ to $t_2$ are computed as
\begin{equation}\label{eq:ir_definition}
    \begin{split}
        IR_{t_1:t_2,k}&= (t_2-t_1+1)\dfrac{\widehat{\mu}_{ret_{t_1:t_2,k}}}{\widehat{\sigma}_{ret_{t_1:t_2,k}}}\\
        &= \frac{\sum\limits_{t=t_1}^{t_{2}} ret_{t,k}}{\sqrt{\frac{1}{t_2 - t_1}\sum\limits_{t=t_1}^{t_2}\left( ret_{t,k} - \frac{1}{t_2 - t_1 + 1}\sum\limits_{t=t_1}^{t_{2}} ret_{t,k} \right)^{2}}}.
    \end{split}
\end{equation}
Finally, the ranking of teams within an interval from $t_1$ to $t_2$ can be computed as
\begin{equation}
    rank_{t_1:t_2,k}=\textrm{card}(\left\{ k'| 1 \leq k' \leq K: IR_{t_1:t_2,k'} \geq IR_{t_1:t_2,k} \right\}).
\end{equation}
The top 5 teams in the global ranking
\begin{equation}
    rank_{T_{1}:T_{12},k} \leq 5
\end{equation}
and the top 3 teams in the quarterly ranking
\begin{equation}
    rank_{T_{(q-1)*3+1}:T_{(q-1)*3+3},k} \leq 3 \qquad q \in \{1,2,3,4\}
\end{equation}
are rewarded.\footnote{
    The official competition rules are available in \citet{makridakisM6FinancialDuathlon2022}.
}

\section{The Role of Luck}\label{section:role_of_luck}
The exceptional investment returns achieved by the top teams in M6 raise a critical question: 
Are these performances indicative of a genuine ability to consistently outperform the market, challenging the Efficient Market Hypothesis (EMH), or could they be explained by the fact that among the large number of teams competing in M6, some are bound to achieve good results purely by chance?
More formally, are the $IR_{T_{1}:T_{12},k}$ observed among the M6 teams compatible with the null hypothesis that the $\mathbb{E}[IR_{T_{1}:T_{12},k}]$ of \emph{all} teams are equal?
The methodology for testing the equality of Sharpe ratios has evolved since the pioneering work of \citet{jobsonPerformanceHypothesisTesting1981} and \citet{memmelPerformanceHypothesisTesting2003}.
\citet{ledoitRobustPerformanceHypothesis2008} generalized the test of equality of expected Sharpe ratios for two portfolio strategies to account for non-normality and time-series autocorrelation.
Particularly relevant to our application, \citet{leungTestingEqualityMultiple2008} propose a test of equality for many portfolio strategies under the assumption of normality and temporal independence.
\citet{wrightTestEqualityMultiple2014} extend the multivariate test to accommodate non-normality and conditional heteroskedasticity.

\citet{wrightTestEqualityMultiple2014} (WYY henceforth) test the null hypothesis\footnote{
    Through this section, we will use the abbreviated notation $IR_{k}$ to denote $IR$ of team $k$ over the whole period $T_{1}:T_{12}$ computed without the log transform in Eq. \ref{eq:log_transform} and without the scaling constant $t_{2}-t_{1}+1$ as it is the definition commonly adopted in the literature regarding the inference about Sharpe ratios.
    Practically speaking however, the differences are negligible.
}
\begin{equation}
    H_{0}: \mathbb{E}[IR_{1}] = \mathbb{E}[IR_{2}] = ... = \mathbb{E}[IR_{K}].
\end{equation}
Let us define
\begin{equation}
    IR = [IR_{1}, IR_{2}, ... , IR_{K}]^{\top}
\end{equation}
and $K-1 \times K$ matrix
\begin{equation}
    Q = \left[ \begin{array}{cccccc}
    1 & -1 & 0 & \cdots & 0 & 0\\
    0 & 1 & -1 & \cdots & 0 & 0\\
    \vdots & \vdots & \vdots & \ddots & \vdots & \vdots \\
    0 & 0 & 0 & \cdots & 1 & -1
    \end{array} \right].
\end{equation}
As shown in \citet{wrightTestEqualityMultiple2014}, one should reject the $H_{0}$ at level $\alpha$ whenever
\begin{equation}
    T^2 = (t_{2}-t_{1}+1)(QIR)^{\top}(Q\hat{\Omega} Q^{\top})^{-1}(QIR) > q_{1-\alpha}(\chi^2_{K-1})
\end{equation}
where $q_{1-\alpha}(\chi^2_{K-1})$ is the $1-\alpha$ quantile of a $\chi^2$ distribution with $K-1$ degrees of freedom and where $\hat{\Omega}$ is a HAC estimator of asymptotic variance of $IR$.\footnote{
    This estimator accounts for the fact that $IR_{k}$ is a nonlinear function of first and second moments $\hat{m}_{1}^{k} = (t_{2}-t_{t}+1)^{-1} \sum_{t=t_{1}}^{t_{2}} RET_{t,k}$ and $\hat{m}_{2}^{k} = (t_{2}-t_{t}+1)^{-1} \sum_{t=t_{1}}^{t_{2}} (RET_{t,k})^2$.
    More details are available in \citet{wrightTestEqualityMultiple2014}.
}

The application of the WYY test to the M6 competition is challenging due to the large number of teams compared to the relatively short time window over which the $IR_{k}$ are measured.
The empirical demonstration in \citet{leungTestingEqualityMultiple2008} and \citet{wrightTestEqualityMultiple2014} involves 18 portfolios over a span of 8 years, which contrasts sharply with the 163 teams who participated during the whole duration of 1 year in M6.
Before applying the test, it is crucial to validate that even in this scenario, the distribution of the $T^{2}$ statistic can be reasonably approximated by its limiting $\chi^{2}_{K-1}$ distribution, ensuring that any rejection is not merely due to finite sample distortions.
To address this, we introduce a highly stylized model of the competition (under the presumption that $H_{0}$ holds) replicating key statistics observed in M6.
By repeatedly simulating this model, we can assess potential level distortions of the WYY test under these extreme conditions.

\subsection{Stylized Model of the Competition}\label{subsection:stylized_model}

We assume that the assets under consideration are homogeneous, with their returns jointly normally distributed and independent over time.
\begin{itemize}
    \item[A1]  \emph{distribution of returns:}
        \begin{equation}
            r_{:,t} \overset{\mathrm{IID}}{\sim} N(\vec{\mu}_{r}, \Sigma_{r})
        \end{equation}
        \emph{where}
        \begin{equation}
            \vec{\mu}_{r}=\mathbbm{1}\mu_{r}
        \end{equation}
        \begin{equation}
            \Sigma_{r}=I\sigma_{r,r} + (J-I)\sigma_{r,r'}
        \end{equation}
\end{itemize}
The assumption of temporal independence appears like a reasonable approximation in light of the EMH \citep[see, e.g.,][]{malkielReflectionsEfficientMarket2005}.\footnote{For simplicity, we disregard the dependence in higher moments, which is well-documented in the literature \citep[see, e.g.,][]{dieboldModelingVolatilityDynamics1995}.}

The value of $\widehat{\mu}_{r}$ corresponds to the long-term average yearly nominal returns of $9.75\%$ \citep{webster500Returns19302023}:
\begin{equation}
    \widehat{\mu}_{r}=0.00037,
\end{equation}
and the covariance matrix is estimated based on observed returns from the duration of the competition:
\begin{equation}
    \widehat{\sigma}_{r,r}=0.00038,
\end{equation}
\begin{equation}
    \widehat{\sigma}_{r,r'}=0.00013.
\end{equation}
More details about the estimation procedure are available in \ref{appendix:returns_distribution}.

To simulate the competition repeatedly, it is necessary to specify how teams choose their portfolios. 
To reduce complexity, we restrict this \emph{baseline portfolio} to the following ternary choice.
\begin{itemize}
    \item[A2]  \emph{baseline portfolio:}
        \begin{equation}\label{eq:baseline_portfolio}
            w_{:,m,k} \overset{\mathrm{w/or}}{\sim} \begin{cases}
                \dfrac{1}{n_{+}+n_{-}} & n_{+1}\, \text{times} \\
                0  & n_{0}\, \text{times}  \\
                \dfrac{-1}{n_{+}+n_{-}} & n_{+1}\, \text{times} \\
            \end{cases}
        \end{equation}
\end{itemize}
The position on each asset is drawn at random\footnote{
    The sampling is performed without replacement to ensure that $\sum_{i=1}^{I}|w_{i,m,k}| \in [0.25,1]$.
}, disregarding past realized returns and all other available information.
This is motivated by the assumption of homogeneity of assets and temporal independence (A1); indeed, there is nothing to be predicted in the first place, by assumption.

The frequency of short, zero, and long positions is estimated by matching moments of the actual investment returns $IR_{T_{m},:}$ observed in the public leaderboard with their simulated counterparts using the method of simulated moments \citep{mcfaddenMethodSimulatedMoments1989}:
\begin{equation}
    \begin{split}
        \hat{n}_{+1}&=38\\
        \hat{n}_{0}&=29\\
        \hat{n}_{-1}&=33.
    \end{split}
\end{equation}
Details of the estimation procedure are available in \ref{appendix:baseline_portfolio}.
From the estimated values, it is apparent that teams, on average, maintain somewhat concentrated portfolios ($\hat{n}_{0} > 0$) and exhibit an affinity towards long positions ($\hat{n}_{+1} > \hat{n}_{-1}$).

\begin{table}[!htbp]
    \fontsize{5}{5}\selectfont
    \centering
    \begin{filecontents*}{leaderboard_comparison.csv}
"obs_month","obs_mean","sim_mean","obs_sd","sim_sd","obs_q_01","sim_q_01","obs_q_99","sim_q_99"
1,1.53,1.47,3.39,3.19,-9.42,-6.04,6.39,8.93
,,(0.25),,(0.21),,(1.01),,(1.05)
2,-3.23,-2.13,4.31,4.22,-9.18,-11.42,9.36,7.76
,,(0.33),,(0.25),,(1.24),,(1.22)
3,0.70,0.55,3.17,3.71,-7.80,-8.17,9.95,9.06
,,(0.29),,(0.23),,(1.15),,(1.08)
4,-2.37,-2.03,3.48,4.13,-10.44,-11.12,5.93,8.03
,,(0.32),,(0.26),,(1.21),,(1.28)
5,-0.28,0.23,2.25,3.34,-5.58,-7.52,4.93,7.95
,,(0.26),,(0.21),,(1.04),,(1.03)
6,2.97,2.02,4.42,4.04,-7.12,-7.37,10.33,11.01
,,(0.31),,(0.24),,(1.14),,(1.13)
7,-2.20,-2.62,4.73,3.51,-8.59,-10.43,11.31,5.73
,,(0.27),,(0.22),,(1.01),,(1.04)
8,-1.37,-2.44,4.59,3.92,-6.94,-11.06,8.46,7.07
,,(0.31),,(0.24),,(1.08),,(1.26)
9,2.74,3.13,5.91,4.50,-10.08,-7.79,9.68,12.56
,,(0.35),,(0.26),,(1.35),,(1.06)
10,-0.48,0.08,3.82,4.93,-10.55,-10.91,8.46,11.53
,,(0.38),,(0.29),,(1.34),,(1.45)
11,0.04,0.22,2.14,3.57,-5.94,-8.13,4.52,8.23
,,(0.28),,(0.22),,(1.03),,(1.00)
12,0.49,1.23,5.68,5.49,-11.78,-11.04,9.10,11.89
,,(0.43),,(0.28),,(1.19),,(0.98)
global,-3.06,0.50,9.43,13.20,-26.27,-29.06,26.22,29.82
,,(1.04),,(0.74),,(3.36),,(3.32)
\end{filecontents*}
\pgfplotstabletypeset[
        col sep = comma,
        ignore chars={"},
        every head row/.style={before row={%
                        \toprule
                        \multicolumn{1}{c}{} & \multicolumn{2}{c}{$ mean(IR_{T_{m},:}) $} &  \multicolumn{2}{c}{$ sd(IR_{T_{m},:}) $} &\multicolumn{2}{c}{$ q_{0.01}(IR_{T_{m},:}) $}  & \multicolumn{2}{c}{$ q_{0.99}(IR_{T_{m},:}) $}\\
                    },after row=\midrule},
        every last row/.style={after row=\bottomrule},
        display columns/0/.style={string type, column name={$m$}},
        display columns/1/.style={string type, column name={obs.}},
        display columns/2/.style={string type, column name={sim.}},
        display columns/3/.style={string type, column name={obs.}},
        display columns/4/.style={string type, column name={sim.}},
        display columns/5/.style={string type, column name={obs.}},
        display columns/6/.style={string type, column name={sim.}},
        display columns/7/.style={string type, column name={obs.}},
        display columns/8/.style={string type, column name={sim.}}
    ]{leaderboard_comparison.csv}
    \caption{
        Observed \& Simulated $IR_{T_{m},:}$\\
        \footnotesize
        Observed mean, standard deviation, $1\%$, and $99\%$ quantiles of $IR_{T_{m},:}$ for individual submissions and globaly, along with the mean of their simulated counterparts under A2.
        Standard deviations across simulations in parentheses.
    }
    \label{tab:leaderboard_comparison}
\end{table}

To verify that the baseline portfolio approximately captures the key aspects of actual teams’ behavior, we repeatedly (10,000 repetitions) simulate the competition with 163 teams following the baseline portfolio and compute the resulting $IR_{T_{m},:}$ using the returns $r_{i,t}$ actually observed during M6.
Table \ref{tab:leaderboard_comparison} displays key descriptive statistics (means, standard deviations, and quantiles) of $IR_{T_{m},:}$ alongside their standard deviations across simulations and compares them with the same statistics observed in the M6 leaderboard. 
Despite the simplicity of the baseline portfolio, the simulated statistics of $IR_{T_{m},:}$ under A2 align remarkably well with those observed in the public leaderboard. 
For submissions in which teams' $IR_{T_{m},:}$ on average declined, the simulated $IR_{T_{m},:}$ under A2 also declines, and vice versa.
More importantly, the dispersion of simulated $IR_{T_{m},:}$ also approximately matches the dispersion of the actually observed $IR_{T_{m},:}$.

\subsection{Equality of Sharpe Ratios}\label{subsection:equality_IR}

To assess the level distortions of the WYY test, we repeatedly simulate the competition under A1 and A2 with $K \in \{5,50,163\}$ teams (1000 repetitions per combination).
Table \ref{tab:test_IR_level_table_asy} shows average rejection rates of the WYY test with asymptotic critical values at levels $\alpha \in \{0.01,0.05,0.1\}$.
While the WYY test exhibits the correct level in situations with $K=5$, it already severely over-rejects at $K=50$.
For $K=163$, it rejects $H_{0}$ 100\% of the time despite $H_{0}$ being satisfied. 
Clearly, the WYY test in its canonical form cannot be readily applied as $T^{2}$ appears to be poorly approximated by its limiting $\chi^{2}_{K-1}$ distribution, given the duration of one year and $K=163$.

\begin{table}[!htbp]
    \centering
    \label{tab:combined_rejection_rates}
    \fontsize{5}{5}\selectfont 
    \begin{subtable}{.5\linewidth} 
        \centering
        \begin{filecontents*}{test_IR_level_table_asy.csv}
"level_num","5_WYY","50_WYY","163_WYY"
0.01,0.012,0.796,1
0.05,0.052,0.898,1
0.1,0.115,0.932,1
\end{filecontents*}
\pgfplotstabletypeset[
        col sep = comma,
        ignore chars={"},
        every head row/.style={before row={\toprule}, after row=\midrule},
        every last row/.style={after row=\bottomrule},
        display columns/0/.style={string type, column name={$\alpha$}},
        display columns/1/.style={dec sep align, precision = 3, fixed, fixed zerofill=true, column name={$K=5$}},
        display columns/2/.style={dec sep align, precision = 3, fixed, fixed zerofill=true, column name={$K=50$}},
        display columns/3/.style={dec sep align, precision = 3, fixed, fixed zerofill=true, column name={$K=163$}}
        ]{test_IR_level_table_asy.csv}
        \caption{WYY (asy.)}
        \label{tab:test_IR_level_table_asy}
    \end{subtable}%
    \begin{subtable}{.5\linewidth}
        \centering
        \begin{filecontents*}{test_IR_level_table_wb.csv}
"level_num","5_WB-WYY","50_WB-WYY","163_WB-WYY"
0.01,0.01,0.014,0.017
0.05,0.051,0.075,0.055
0.1,0.097,0.12,0.105
\end{filecontents*}
\pgfplotstabletypeset[
        col sep = comma,
        ignore chars={"},
        every head row/.style={before row={\toprule}, after row=\midrule},
        every last row/.style={after row=\bottomrule},
        display columns/0/.style={string type, column name={$\alpha$}},
        display columns/1/.style={dec sep align, precision = 3, fixed, fixed zerofill=true, column name={$K=5$}},
        display columns/2/.style={dec sep align, precision = 3, fixed, fixed zerofill=true, column name={$K=50$}},
        display columns/3/.style={dec sep align, precision = 3, fixed, fixed zerofill=true, column name={$K=163$}}
        ]{test_IR_level_table_wb.csv}
        \caption{WYY (w.b.)}
        \label{tab:test_IR_level_table_wb}
    \end{subtable}
    \caption{Rejection Rates of WYY Tests\\
    Rejection rates of the WYY test with a) asymptotic critical values and b) bootstrapped critical values under different numbers of teams $K$ and levels $\alpha$.
    }
\end{table}

To rectify these finite sample level distortions, we obtain critical values via null-imposed bootstrap \citep[see][section 1.8]{politisSubsampling1999}, as mentioned in \citet{ledoitRobustPerformanceHypothesis2008}.
Specifically, we perform a variant of wild bootstrap where the resampled portfolio weights $\tilde{w}_{i,m,k}$ are defined as:
\begin{equation}
    \tilde{w}_{i,m,k}=\xi_{m,k} w_{i,m,k} \quad \xi_{m,k}\stackrel{\text{IID}}{\sim}
\begin{cases}
    \phantom{-}1 & \text{w.p. } 0.5 \\
    -1 & \text{w.p. } 0.5,
\end{cases}
\end{equation}
and where $\tilde{IR}_{k}$ and $\tilde{T}^{2}$ are the corresponding Sharpe ratio under the resampled weights and the overall testing statistic computed using $\tilde{IR}_{k}$, respectively.
This leads to the rejection rule
\begin{equation}
    T^2 = (t_{2}-t_{1}+1)(QIR)^{\top}(Q\hat{\Omega} Q^{\top})^{-1}(QIR) > q_{1-\alpha}(\tilde{T}^{2})
\end{equation}
where $q_{1-\alpha}(\tilde{T}^{2})$ is the $1-\alpha$ quantile of $\tilde{T}^{2}$ across bootstrap iterations.
As $\tilde{IR}_{k}$ is symmetrically distributed around $0$ by construction, it follows that $\mathbb{E}[\tilde{IR}_{k}]=0$ and that $H_{0}$ is satisfied.
Consequently, following \citet{wrightTestEqualityMultiple2014}, 
\begin{equation}
    \tilde{T}^{2} \stackrel{\text{d}}{\rightarrow} \chi_{K-1}^{2},
\end{equation}
implying that such an approach is asymptotically valid.
To explore its small sample size properties, we repeat the simulation experiment as with the WYY test with analytical critical values.
As shown in Table \ref{tab:test_IR_level_table_wb}, rejection rates of the WYY test with bootstrap critical values (1000 bootstrap repetitions) are much closer to their nominal values, even with $K$ as high as $163$.

Equipped with a correctly leveled test, we can proceed with addressing the main question.
Table \ref{tab:test_IR_pvalue_table} shows the p-value of the WYY test with bootstrap critical values (1000 bootstrap repetitions) applied to portfolio weights submitted in M6.\footnote{
    14 teams whose submissions were identical to the \emph{M6 dummy} were counted as a single team to ensure that the covariance matrix is of full rank.
} 
The observed $\{IR_{k}\}_{k=1}^{K}$ appear to be compatible with the null hypothesis that $\mathbb{E}[IR_{k}]$ are equal across teams and, in particular, equal to the benchmark equal-weighted long portfolio \emph{M6 dummy}.

\begin{table}[!htbp]
    \fontsize{5}{5}\selectfont
    \centering
    \begin{filecontents*}{test_IR_pvalue_table.csv}
"test","pvalue"
"WYY (w.b.)",0.926
\end{filecontents*}
\pgfplotstabletypeset[
    col sep = comma,
    ignore chars={"},
    every head row/.style={before row={%
    \toprule
    },after row=\midrule},
    every last row/.style={after row=\bottomrule},
    display columns/0/.style={string type, column name={test}},
    display columns/1/.style={dec sep align, precision = 3, fixed, fixed zerofill=true, column name={$p$-value}},
    ]{test_IR_pvalue_table.csv}
    \caption{Test of $H_{0}:\forall k, 1\leq k\leq K: \mathbb{E}[IR_{k}]=const.$\\
        P-value of WYY test with bootstrap critical values (1,000 repetitions) applied to $\{IR_{k}\}_{k=1}^{K}$ observed in M6.
    }
    \label{tab:test_IR_pvalue_table}
\end{table}

This conclusion also seems to be supported directly by Table \ref{tab:leaderboard_comparison}, as simulated $IR_{T_{m},:}$ under A2 mimics not only means and dispersions of $IR_{T_{m},:}$ observed among teams but also its tail behavior.
For instance, the $99\%$ quantile of $IR_{T_{1}:T_{12},:}$ if teams simply submitted their portfolios at random according to A2 is $29.82$ with a standard deviation of $3.32$.
This indeed encompasses the actual observed $99\%$ quantile of the total $IR_{T_{1}:T_{12},:}$; $26.22$, confirming that it is possible to obtain as extreme $IR_{T_{1}:T_{12},:}$ as those observed in the competition, even if $H_{0}$ is satisfied.

\subsection{Sharpe Ratios and Rank}\label{subsection:IR_and_Rank}

In addition to testing $H_{0}$, it is interesting to consider a related question regarding the role of luck in the competition: What degree of predictability of returns would be needed to reliably secure a top rank in the competition?
To answer this, we slightly extend A1 to allow for semi-strong efficiency of markets with a degree of predictability $\lambda$.
\begin{itemize}
    \item[A1']  \emph{distribution of returns:}\\
        \emph{
            On top of A1, returns can be further decomposed into two independent components: unpredictable component $r_{:,t}^{u}$ and predictable component $r_{:,t}^{p}$:
        }
        \begin{equation}
            r_{:,t}=r_{:,t}^{u}+r_{:,t}^{p} \qquad
            \begin{array}{l}
                r_{:,t}^{u} \overset{\mathrm{IID}}{\sim} N((1-\lambda)\vec{\mu}_{r}, (1-\lambda)\Sigma_{r}) \\
                r_{:,t}^{p} \overset{\mathrm{IID}}{\sim} N(\lambda\vec{\mu}_{r}, \lambda\Sigma_{r}).
            \end{array}
        \end{equation}
\end{itemize}
This allows us to assess the performance of a team who can predict returns to the degree $\lambda$, 
For this type of team, the tangency portfolio that maximizes expected risk-adjusted returns \cite[see e.g.,][]{kourtisSharpeRatioEstimated2016} is:
\begin{equation}\label{eq:tangency_portfolio}
    w_{:,m,k}=\Sigma_{r}^{-1}((1-\lambda) \vec{\mu}_{r} + \sum_{t\in T_{m}}r_{:,t}^{p}).
\end{equation}

We repeatedly (100,000 repetitions) simulate the stylized model of the competition under A1' with 162 teams following the baseline portfolio (A2) and one team following the tangency portfolio in Eq. \ref{eq:tangency_portfolio} with varying degrees of predictability $\lambda$.
Table \ref{tab:state_table_simulated_tangency} displays the mean attained returns $\mathbb{E}[IR_{T_{1}:T_{12},k}]$ and the probability of securing at least 1st, 5th, 10th, and 20th rank in the competition for individual portfolio strategies. 
In an environment with a dispersion of $IR_{T_{1}:T_{12},:}$ comparable to M6, a team capable of consistently attaining almost double the market returns (relative to the performance of the tangency portfolio with $\lambda = 0$, which collapses to the equal-weighted long portfolio) can still secure up to 10th place only with a probability of $0.146$. 
This demonstrates that, even with substantially better $\mathbb{E}[IR_{T_{1}:T_{12},k}]$, the outcome in terms of top rank is still far from certain and is still significantly influenced by luck, even when measured over the span of one year.

\begin{table}[!htbp]
    \fontsize{5}{5}\selectfont
    \centering
    \begin{filecontents*}{state_table_simulated_tangency.csv}
"portfolio","IR","top_1","top_5","top_10","top_20"
"baseline",2.95839625459523,0.00599,0.03057,0.06145,0.12251
"tangency ($\lambda$ = 0)",6.37471247914413,7e-04,0.00741,0.02404,0.0791
"tangency ($\lambda$ = 0.0001)",8.98072394414767,0.0043,0.03059,0.07296,0.17043
"tangency ($\lambda$ = 0.0003)",11.760446301817,0.01536,0.07577,0.14628,0.26892
"tangency ($\lambda$ = 0.001)",18.2857137595245,0.06802,0.20822,0.31708,0.46421
\end{filecontents*}
\pgfplotstabletypeset[
    col sep = comma,
    ignore chars={"},
    every head row/.style={before row={%
    \toprule
    \multicolumn{11}{c}{\hspace{120pt} $ P(rank_{T_{1}:T_{12},k}\leq q)  $}\\
    },after row=\midrule},
    every last row/.style={after row=\bottomrule},
    display columns/0/.style={string type, column name={portfolio}},
    display columns/1/.style={dec sep align, precision = 2, fixed, fixed zerofill=true, column name={$\mathbb{E}[IR_{T_{1}:T_{12},k}]$}},
    display columns/2/.style={dec sep align, precision = 3, fixed, fixed zerofill=true, column name={$ q = 1$}},
    display columns/3/.style={dec sep align, precision = 3, fixed, fixed zerofill=true, column name={$ q = 5$}},
    display columns/4/.style={dec sep align, precision = 3, fixed, fixed zerofill=true, column name={$ q = 10$}},
    display columns/5/.style={dec sep align, precision = 3, fixed, fixed zerofill=true, column name={$ q = 20$}}
    ]{state_table_simulated_tangency.csv}
    \caption{Relationship between $\mathbb{E}[IR_{T_{1}:T_{12},k}]$ and Rank}
    \label{tab:state_table_simulated_tangency}
\end{table}

Another interesting observation is that the relationship between $\mathbb{E}[IR_{T_{1}:T_{12},k}]$ and $ P(rank_{T_{1}:T_{12},k}\leq q) $ does not appear to be monotone.
The baseline portfolio with expected returns of $\mathbb{E}[IR_{T_{1}:T_{12},k}]=2.96$ has a higher probability of securing a top rank than the tangency portfolio with $\lambda = 0$, which attains $\mathbb{E}[IR_{T_{1}:T_{12},k}]=6.37$.
This puzzling result highlights that the task of securing a good rank in the competition might not necessarily be identical to that of attaining the best investment returns in expectation, something which will be examined in more detail in the following section.

\section{The Role of Strategic Considerations}\label{section:role_of_strategic}

Teams in the investment challenge could theoretically improve their chances of securing a top rank through several methods, beyond the challenging task of increasing their expected risk-adjusted returns, $\mathbb{E}[IR_{T_{1}:T_{12},k}]$.
Directly manipulating portfolio volatility to increase (resp. decrease) the probability of achieving (resp. losing) a sufficiently good rank \citep[see, e.g.,][]{brownTournamentsTemptationsAnalysis1996,eltonIncentiveFeesMutual2003} is not directly applicable to the M6 due to the use of risk-adjusted returns as the evaluation metric.
Altering the probability of securing the top rank by including assets with different degrees of skewness in the portfolio is also documented primarily in environments where plain returns are used as the evaluation metric \citep[see, e.g.,][]{linFundConvexityTail2011,dijkRankMattersImpact2014}.
While this strategy is theoretically applicable, we did not observe top-performing teams taking disproportionately long positions on assets with positively skewed returns or short positions on assets with negatively skewed returns (see Figure \ref{fig:submission_analysis_rank_skew} in \ref{appendix:auxiliary_results}), suggesting that this channel was likely of limited importance.

Instead, we focus on the importance of the correlation of returns from different investment strategies \citep[see, e.g.,][]{niekenRisktakingTournamentsTheory2010,krasnyAssetPricingStatus2011}, which can be, to some extent, regulated by the proportion of long vs. short positions in one's portfolio, as teams in M6 predominantly opted for long positions.
To explore how such a strategic approach might look and to what degree it might be beneficial, we derive the optimal strategy for maximizing the probability of securing the desired rank in a stylized environment under assumptions A1 and A2.

\subsection{Optimal Strategic Portfolio}\label{subsection:strategic_portfolio}

The conventional way of approaching the competition is to maximize the stated objective:
\begin{equation}\label{eq:conventional_objective}
    \underset{ w_{:,m,k}}{\textrm{max}}\, \mathbb{E}\left[ IR_{T_m,k} | \mathcal{S}_{m-1}\right] \qquad 1\leq m\leq 12,
\end{equation}
where $\mathcal{S}_{m}$ denotes the state at submission $m$, containing all the relevant information available at that time.
To account for the adversarial nature of the competition, teams might focus directly on optimizing the probability of securing at least the $q$-th rank on the global leaderboard:
\begin{equation}
    P\left(rank_{T_1:T_{12},k}\leq q\right).
\end{equation}
This problem differs from the one presented in Eq. \ref{eq:conventional_objective} as it cannot be separated into $12$ independent optimization sub-problems.
The decision regarding portfolio weights $w_{:,m,k}$ depends crucially on the current leaderboard, especially the current rank $rank_{T_1:T_{m-1},k}$, which is encompassed in $\mathcal{S}_{m-1}$.
Likewise, when choosing portfolio weights $w_{:,m,k}$, one must consider its effects on the probability of securing at least the $q$-th rank on the global leaderboard, while also accounting for the fact that it is possible to further alter the ranking in submissions yet to come.

The value function measuring the probability of securing at least the $q$-th rank associated with the dynamic programming problem described above is
\begin{equation}
    \begin{split}
        &V_{m}(\mathcal{\mathcal{S}}_{m-1})=\\
        &\begin{cases}
            \underset{ w_{:,m,k}}{\textrm{max}} \, V_{m+1}(\mathcal{\mathcal{S}}_{m})P\left( \mathcal{\mathcal{S}}_{m}| \mathcal{\mathcal{S}}_{m-1}, w_{:,m,k}\right) & 1\leq m \leq 11 \\
            \underset{ w_{:,m,k}}{\textrm{max}} \, P\left(rank_{T_1:T_{12},k}\leq q| \mathcal{\mathcal{S}}_{m-1}, w_{:,m,k}\right)                                    & m=12.           \\
        \end{cases}
    \end{split}
\end{equation}
To solve this optimization, we can proceed with backward induction, repeatedly simulating the competition under assumptions A1 and A2 and numerically solving the optimization sub-problem for $m=12$ with various $\mathcal{S}_{11}$.
Then, armed with the knowledge of $V_{12}(\mathcal{S}_{11})$, we can continue solving the sub-problem for $m=11$ and so forth.

To reduce computational complexity, we restrict $\mathcal{S}_{m}$ to contain only the distance to the $q$-th largest value of $IR_{T_{1}:T_{m},k}$ among the competitors:
\begin{equation}
    \mathcal{S}_m = \lbrace \Delta_{m} \rbrace \qquad \Delta_{m} = IR_{T_1:T_m,k} - \underset{k' \neq k}{\textrm{max-$q$-th}} \,IR_{T_1:T_m,k}.
\end{equation}
Let us denote the proportion of long (resp. short) positions in the portfolio for team $k$ at submission $m$ as
\begin{equation}
    \beta^{+}_{m,k} = \dfrac{\sum_{i=1}^{I}\textrm{max}(w_{i,m,k},0)}{\sum_{i=1}^{I}|w_{i,m,k}|},
\end{equation}
and its complement
\begin{equation}
    \beta^{-}_{m,k} = \dfrac{\sum_{i=1}^{I}\textrm{max}(-w_{i,m,k},0)}{\sum_{i=1}^{I}|w_{i,m,k}|} = 1- \beta^{+}_{m,k}.
\end{equation}
Similarly as in the case of the benchmark portfolio, we restrict the portfolio weights $w_{i,m,k}$ to be sampled without replacement from
\begin{equation}\label{eq:rank_opt_portfolio}
    w_{:,m,k} \overset{\mathrm{w/or}}{\sim} \begin{cases}
        \phantom{+}0.01 & \beta^{+}(\Delta_{m},m)*I\, \text{ times} \\
        -0.01  & \beta^{-}(\Delta_{m},m)*I\, \text{ times}. \\
    \end{cases}
\end{equation}
The crucial difference here is that $\beta^{+}(\Delta_{m},m)$ is now a function of the distance to the $q$-th ranking team and the current submission $m$, thus allowing one to strategically alter the proportion of short positions in the portfolio to improve one's probability of attaining the desired rank.\footnote{
    We consider only positions $-0.01$ and $0.01$ as the presence of zero positions seems to alter the distribution of $IR_{T_{m},k}$, and even more importantly, the distribution of $IR_{T_{m},k}$ relative to $IR_{T_{m},k'}$ of the baseline portfolio, only very little, due to the normalization in Eq. \ref{eq:ir_definition} (see Fig. \ref{fig:position_effects} in \ref{appendix:auxiliary_results}).
}

\begin{figure}
    \centering
    \includegraphics[width=1.0\linewidth]{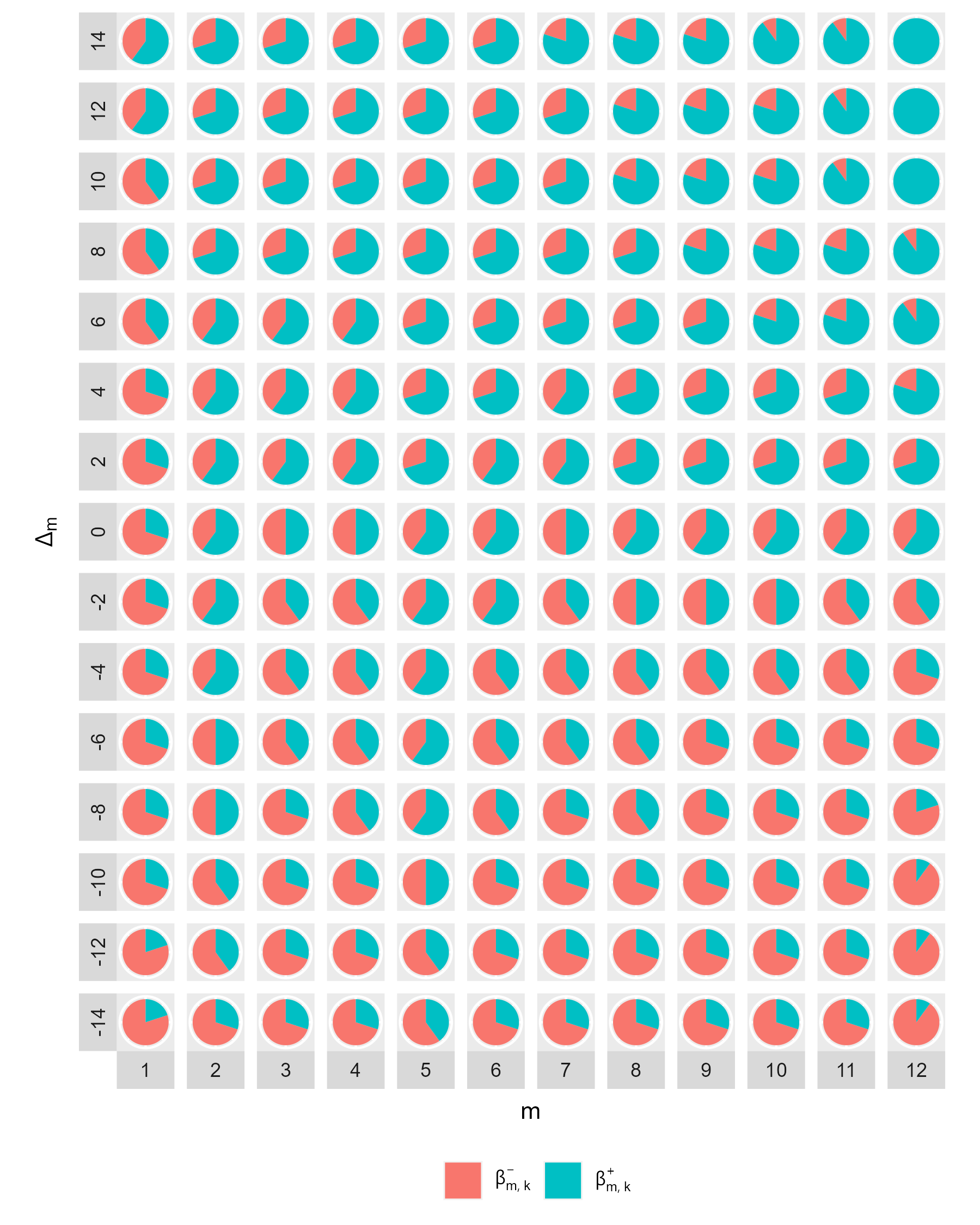}
    \caption{Rank Optimization Portfolio ($q=1$)\\
        \footnotesize
    }
    \label{fig:strategy_diagram_q_1}
\end{figure}

Figure \ref{fig:strategy_diagram_q_1} displays the optimal portfolio for $q=1$ as a function of $m$ and $\Delta_{m}$ computed under A1-A2.\footnote{
    For the purpose of deriving the optimal strategy, we assume that returns are standardized by an estimate of the standard deviation for the respective submission, rather than the estimate over the entire evaluated period.
    This alters the $IR_{T_{1}:T_{12}}$ only minimally and makes the investment returns additive ($IR_{T_{1}:T_{12},k}= \sum_{m=1}^{12}IR_{T_{m,k}}$), which significantly simplifies the dynamic programming problem.
    To further reduce computational requirements, we also restrict $\beta^{+}(\Delta_{m},m)$ to be a multiple of $0.1$.
}
The primary characteristic of the optimal strategy lies in its liberal use of short positions.
In the initial submission, it is optimal to primarily take short positions ($\beta^{+}<0.5$) to create a dispersion between one's own $IR_{T_{1},k}$ and the $IR_{T_{1},k'}$ of competitors who, given A2, submit predominantly long portfolios.
Then, if the current $IR_{T_{1}:T_{m},k}$ is higher or at least comparable to the $IR_{T_{1}:T_{m},k'}$ of the $q$-th ranking competitor, it is optimal to adopt predominantly long positions to mimic the behavior of possible contenders and hence minimize the probability of losing an already sufficiently good rank.
If the $q$-th ranking competitor is in the lead, and the gap $\Delta_{m}$ is either so large or the end of the competition is so close that securing the $q$-th rank is unlikely, it is optimal to adopt more short positions to maximize the dispersion of the difference between one's $IR_{T_{m},k}$ and that of the incumbent (see Fig. \ref{fig:position_effects} in \ref{appendix:auxiliary_results}).
This type of adversarial portfolio, combined with teams' inclination towards long positions, enables one to climb the leaderboard from positions that would otherwise be unsalvageable and to minimize the probability of losing a sufficiently good rank once it has been secured.
The optimal strategy for $q=20$ is similar (see Fig. \ref{fig:strategy_diagram_q_20} in \ref{appendix:auxiliary_results}), except for the less aggressive shorting in the initial submission.

It is important to highlight that the derivation of such a strategy does not necessarily rely on information that would be in principle unavailable to teams at the time of the competition.
In A1, it leverages long-term stock returns averages and a very simple covariance estimate.
In A2, the proclivity of teams to take long positions ($\hat{n}_{+1}>\hat{n}_{-1}$), which lies at the heart of the strategy, can be deduced from the public leaderboard after the end of the first submission $m=1$.
Strategies like this were hence in principle feasible for the teams.
Furthermore, given the very simplifying assumptions made to derive this strategy, it is possible that teams might utilize more complex and nuanced strategies with even better performance.

\subsection{Simulation Results}

To assess the performance of the rank optimization portfolio, we utilize two distinct evaluation environments, each with its specific drawbacks and advantages.
One way of assessing the performance is to repeatedly simulate the stylized model of the competition under A1' and A2 with $K=163$ teams, similarly as in Sub-Section \ref{subsection:IR_and_Rank}.
Out of these, $162$ utilize the baseline portfolio to mimic the typical risk profiles and affinity towards long positions observed in the competition.
The remaining team uses either the rank optimization portfolio with $q \in \{1, 20\}$, the tangency portfolio with/without insider information ($\lambda \in \{0.0003, 0\}$), or the baseline portfolio for comparison.

To ensure robustness with respect to the violation of assumptions A1 and A2, we also re-simulate the competition by bootstrapping the actual observed stock prices $S_{:,t}$ and portfolio weights $w_{:,m,k}$ submitted by individual teams.\footnote{
    See \url{https://github.com/Mcompetitions/M6-methods}.
}
To do so, we resample returns at a 4-week interval frequency such that in each bootstrap iteration, teams face a distinct combination of submission intervals $m$ (sampled with replacement).
We set $K=163$, out of which $162$ teams utilize a bootstrapped portfolio.
The portfolio weights $w_{:,m,k}$ of this bootstrapped portfolio are drawn randomly (with replacement) from the $w_{:,m,k}$ submitted by teams for the corresponding submission interval $m$.
The remaining team uses the rank optimization portfolio with $q\in\{1,20\}$ or the bootstrapped portfolio for comparison.
This approach allows us to completely dispense with assumptions A1 and A2 when evaluating the performance of the rank optimization strategy.
However, the drawback is that the 12 4-week submission intervals are unlikely to be representative of the distribution of returns at large.
As a result, the obtained success rates might be affected by idiosyncrasies observed during this period, and the estimates of $\mathbb{E}[IR_{T_{1}:T_{12},k}]$ are likely to be unreliable.

Table \ref{tab:state_table} displays $\mathbb{E}[IR_{T_{1}:T_{12},k}]$ and the probability of securing at least rank $q$ for $q \in \{1,\,5,\,10,\,20\}$ when simulating the competition under A1' and A2.
The rank optimization portfolio optimized for $q=1$ (resp. $q=20$) secures the 1st (resp. 20th) rank with a probability of $0.019$ (resp. $0.144$), which is better than what one would expect by chance.
The benefit of acting strategically is especially pronounced when focusing on the very top rank ($q=1$).
Here, the probability of securing the desired rank $q$ when specially optimizing for it is comparable to that of the tangency portfolio with $\lambda=0.0003$, which is capable of consistently generating almost double the market returns.\footnote{
    Relative to the $\mathbb{E}[IR_{T_{1}:T_{12},k}]$ of the equal-weighted long portfolio tangency with $\lambda = 0$ that serves as a proxy for market returns.
}
The benefit is less substantial for $q=20$, indicating that the importance of strategic considerations decreases for less extreme ranks.

\begin{table}[!htbp]
    \fontsize{5}{5}\selectfont
    \centering
    \begin{filecontents*}{state_table_simulated_rank_optimization.csv}
"portfolio","IR","beta","top_1","top_5","top_10","top_20"
"baseline",2.95839625459523,0.535211267605033,0.00599,0.03057,0.06145,0.12251
"tangency ($\lambda$ = 0)",6.37471247914413,1,7e-04,0.00741,0.02404,0.0791
"tangency ($\lambda$ = 0.0003)",11.760446301817,0.56893057043435,0.01536,0.07577,0.14628,0.26892
"rank opt. (q = 1)",-6.77784425272037,0.290671750000211,0.01949,0.06258,0.08815,0.12566
"rank opt. (q = 20)",-2.68614826165166,0.438029749999993,0.00483,0.02611,0.05743,0.14449
\end{filecontents*}
\pgfplotstabletypeset[
    col sep = comma,
    ignore chars={"},
    every head row/.style={before row={%
    \toprule
    \multicolumn{11}{c}{\hspace{142pt} $ P(rank_{T_{1}:T_{12},k}\leq q)  $}\\
    },after row=\midrule},
    every last row/.style={after row=\bottomrule},
    display columns/0/.style={string type, column name={portfolio}},
    display columns/1/.style={dec sep align, precision = 2, fixed, fixed zerofill=true, column name={$\mathbb{E}[IR_{T_{1}:T_{12}}]$}},
    display columns/2/.style={dec sep align, precision = 3, fixed, fixed zerofill=true, column name={$\mathbb{E}[\beta^{+}_{m}]$}},
    display columns/3/.style={dec sep align, precision = 3, fixed, fixed zerofill=true, column name={$ q = 1$}},
    display columns/4/.style={dec sep align, precision = 3, fixed, fixed zerofill=true, column name={$ q = 5$}},
    display columns/5/.style={dec sep align, precision = 3, fixed, fixed zerofill=true, column name={$ q = 10$}},
    display columns/6/.style={dec sep align, precision = 3, fixed, fixed zerofill=true, column name={$ q = 20$}}
    ]{state_table_simulated_rank_optimization.csv}
    \caption{Comparison of Performance (Simulated)}
    \label{tab:state_table}
\end{table}

\begin{table}[!htbp]
    \fontsize{5}{5}\selectfont
    \centering
    \begin{filecontents*}{state_table_bootstrapped.csv}
"portfolio","IR","beta","top_1","top_5","top_10","top_20"
"bootstrapped",-2.50850487633106,0.744276143637984,0.00578,0.02985,0.05956,0.12152
"rank opt. (q = 1)",-1.44376047824045,0.313146416666785,0.05884,0.16598,0.22125,0.28661
"rank opt. (q = 20)",-1.03229755284321,0.470836166666664,0.00654,0.04644,0.10997,0.27263
\end{filecontents*}
\pgfplotstabletypeset[
    col sep = comma,
    ignore chars={"},
    every head row/.style={before row={%
    \toprule
    \multicolumn{11}{c}{\hspace{130pt} $ P(rank_{T_{1}:T_{12},k}\leq q)  $}\\
    },after row=\midrule},
    every last row/.style={after row=\bottomrule},
    display columns/0/.style={string type, column name={portfolio}},
    display columns/1/.style={dec sep align, precision = 2, fixed, fixed zerofill=true, column name={$\mathbb{E}[IR_{T_{1}:T_{12},k}]$}},
    display columns/2/.style={dec sep align, precision = 3, fixed, fixed zerofill=true, column name={$\mathbb{E}[\beta^{+}_{m,k}]$}},
    display columns/3/.style={dec sep align, precision = 3, fixed, fixed zerofill=true, column name={$ q = 1$}},
    display columns/4/.style={dec sep align, precision = 3, fixed, fixed zerofill=true, column name={$ q = 5$}},
    display columns/5/.style={dec sep align, precision = 3, fixed, fixed zerofill=true, column name={$ q = 10$}},
    display columns/6/.style={dec sep align, precision = 3, fixed, fixed zerofill=true, column name={$ q = 20$}}
    ]{state_table_bootstrapped.csv}
    \caption{Comparison of Performance (Bootstrapped)}
    \label{tab:state_table_bootstrap}
\end{table}

Table \ref{tab:state_table_bootstrap} displays $\mathbb{E}[IR_{T_{1}:T_{12},k}]$ and the probability of securing rank $q$ in the bootstrap environment, further corroborating these results.
In this case, the rank optimization portfolio optimized for $q=1$ (resp. $q=20$) secures the 1st (resp. 20th) rank with a probability $0.059$ (resp. $0.273$).
These higher success rates compared to the simulated environment are likely attributed to the below-average returns over the duration of M6 relative to the long-term mean returns of $9.75\%$ used in assumption A1'.

\begin{figure}[!htbp]
    \centering
    \includegraphics[width=1\linewidth]{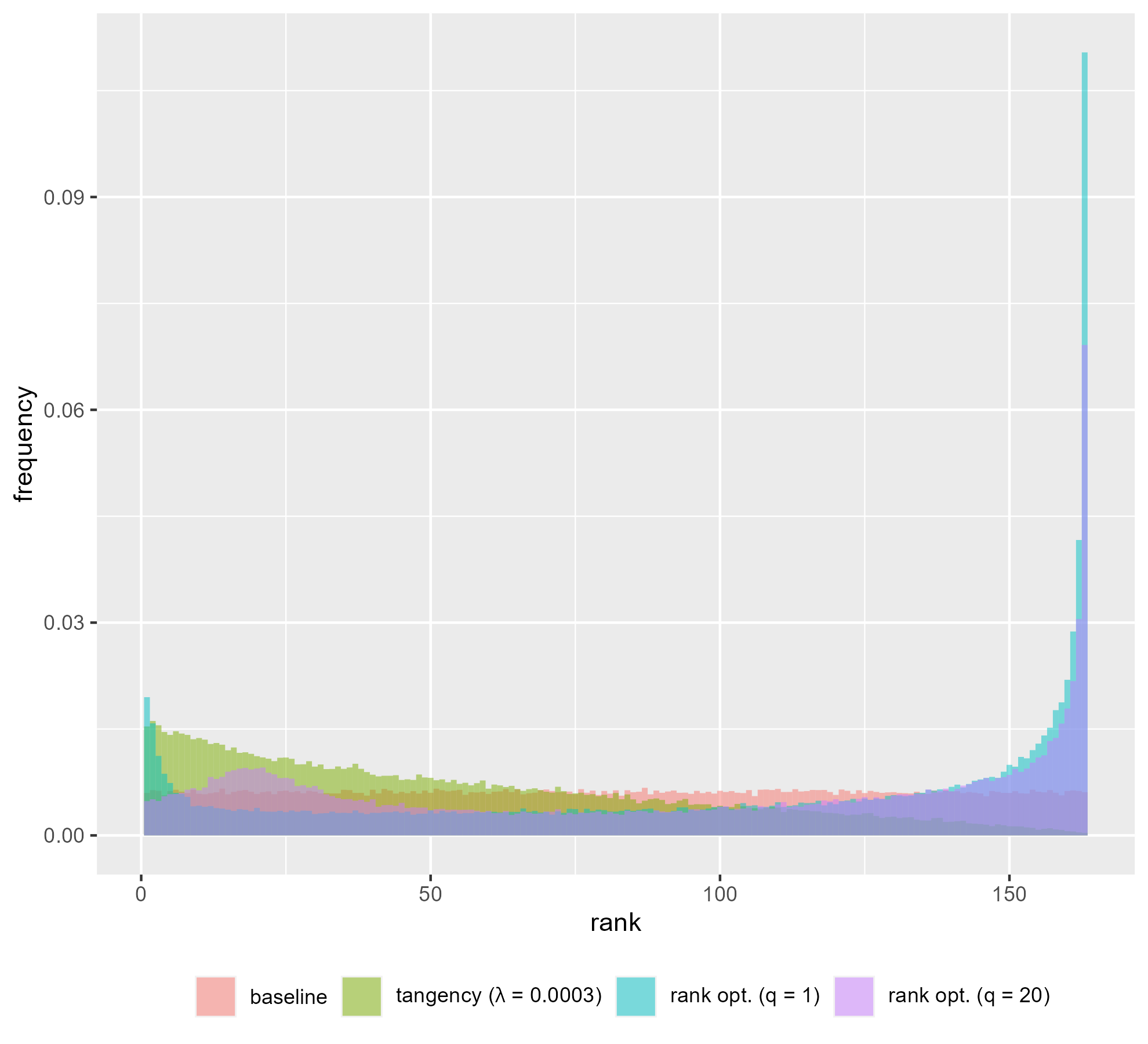}
    \caption{Histogram of Ranks (Simulated)}
    \label{fig:rank_histogram}
\end{figure}

\begin{figure}[!htbp]
    \centering
    \includegraphics[width=1\linewidth]{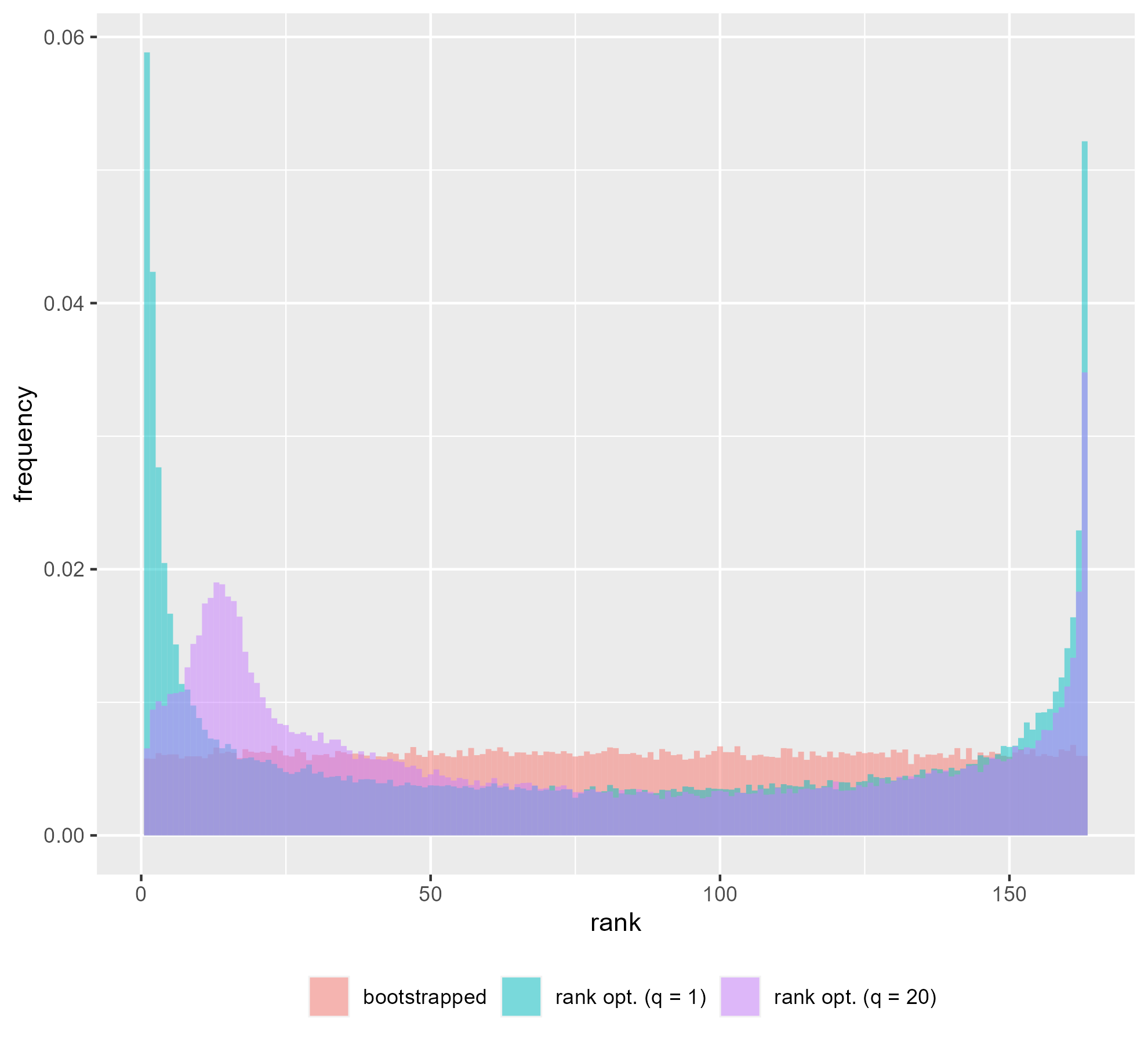}
    \caption{Histogram of Ranks (Bootstrapped)}
    \label{fig:rank_histogram_bootstrap}
\end{figure}

In both Table \ref{tab:state_table} and \ref{tab:state_table_bootstrap}, rank optimization portfolios outperform the baseline/bootstrapped portfolios despite that their $\mathbb{E}[IR_{T_{1}:T_{12},k}]$ is in fact inferior.
Figures \ref{fig:rank_histogram} and \ref{fig:rank_histogram_bootstrap} examine this paradox by plotting the histogram of ranks of individual portfolios in the simulated and bootstrapped environments.
For the tangency portfolio with $\lambda=0.0003$, the superior performance is reflected by increasing the probability of a good rank while at the same time decreasing the probability of a poor rank, as one would expect.
For the rank optimization portfolios, however, the gain in the probability of securing a top rank is achieved \emph{at the cost} of a disproportionately high probability of ending at the very bottom of the leaderboard.
Indeed, there is no free lunch to be found when focusing directly on rank; one may merely alter the tail behavior of $IR_{T_{1}:T_{12},k}$ to make the probability of a spectacular success and a catastrophic failure simultaneously larger.

\subsection{Empirical Evidence}

Disregarding the intricacies of timing the short positions depending on the current ranking, the derived rank optimization strategies can be broadly characterized by their on average higher use of short positions, as demonstrated by column $\mathbb{E}[\beta_{m,k}^{+}]$ in Tables \ref{tab:state_table} and \ref{tab:state_table_bootstrap}.
Indeed, to increase the probability of securing good rank, a team must differentiate one's submission from the predominantly long positions submitted by other teams.
Considering this, it is interesting to analyze directly the submissions of individual teams to see whether the submissions of top-performing teams in M6 indeed exhibit lower $\beta^{+}_{m,k}$ relative to other teams.

\begin{figure}[!htbp]
    \centering
    \includegraphics[width=0.99\linewidth]{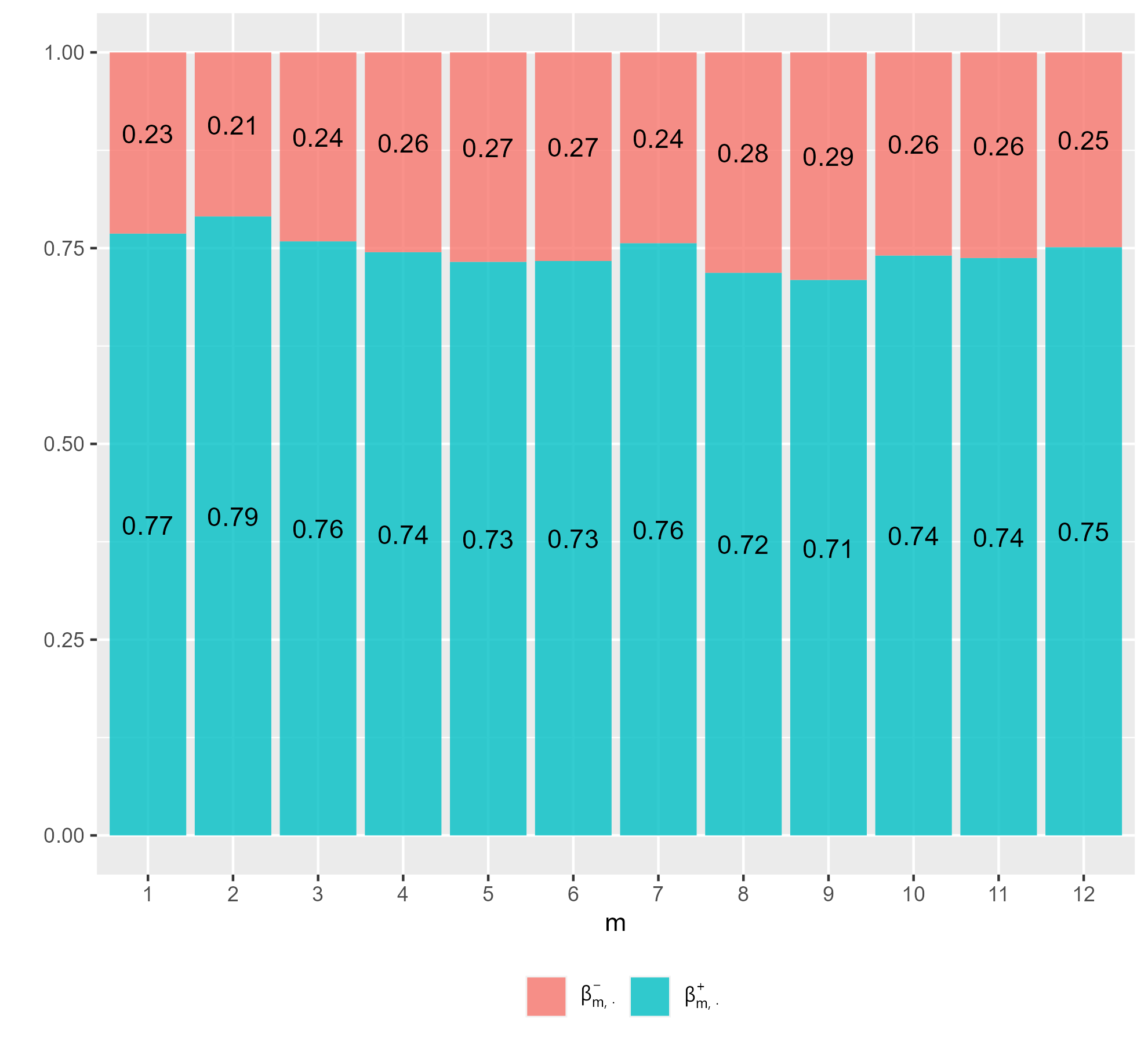}
    \caption{Average Proportion of Long Positions $\beta_{m,k}^{+}$}
    \label{fig:submission_analysis_long}
\end{figure}

\begin{figure}[!htbp]
    \centering
    \includegraphics[width=0.99\linewidth]{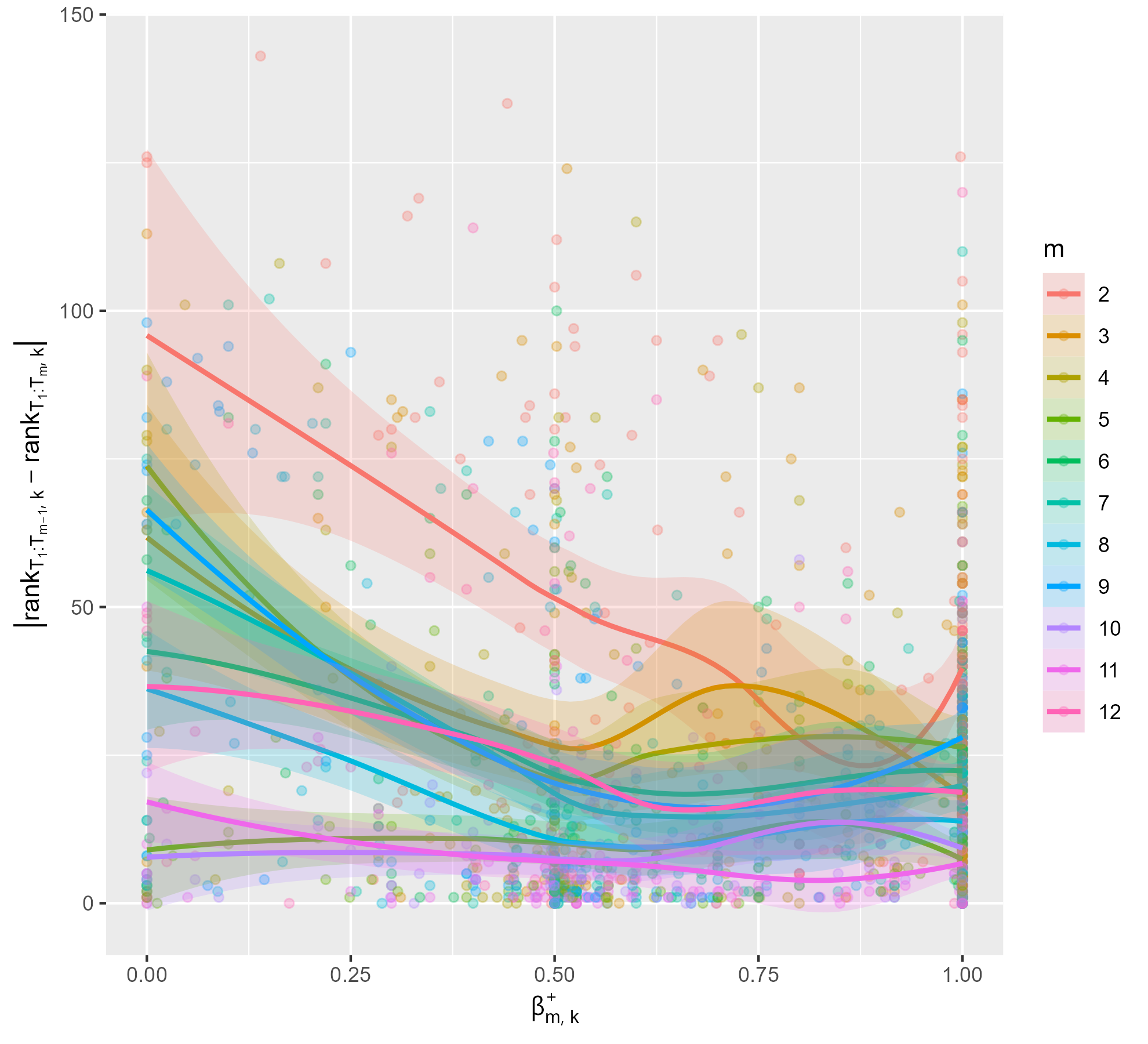}
    \caption{Absolute Change of Rank as Function of $\beta_{m,k}^{+}$\\
        The banded lines represent loess estimates of mean with their corresponding standard deviations.
    }
    \label{fig:submission_analysis_long_rank_diff}
\end{figure}

First, we examine the submitted portfolio weights to verify that they indeed resemble the simulation environment of A1 and A2 used in the theoretical part of this section.
Figure \ref{fig:submission_analysis_long} displays the average $\beta^{+}_{m,:}$ and $\beta^{-}_{m,:}$ across participants.
Teams exhibit a strong preference towards long positions on average, confirming the results of the previous estimation in A2 based solely on the public leaderboard.
This preference appears to be constant in $m$, with approximately 75\% of portfolio positions being long.
Figure \ref{fig:submission_analysis_long_rank_diff} plots teams' absolute changes of rank $|rank_{T_{1}:T_{m-1},k} -rank_{T_{1}:T_{m},k}|$ as function of their $\beta^{+}_{m,k}$.
Low $\beta^{+}_{m,k}$ tend to be followed by large changes of rank, confirming the mechanics underlying the advantage of forming portfolios with low $\beta^{+}_{m,k}$ described in Sub-Section \ref{subsection:strategic_portfolio}.
This relationship is stronger for earlier submissions, which is a consequence of the fact that at later stages of the competition, the $IR_{T_{m},k}$ is computed over longer periods making it less susceptible to change.

\begin{figure}[!htbp]
    \centering
    \includegraphics[width=0.99\linewidth]{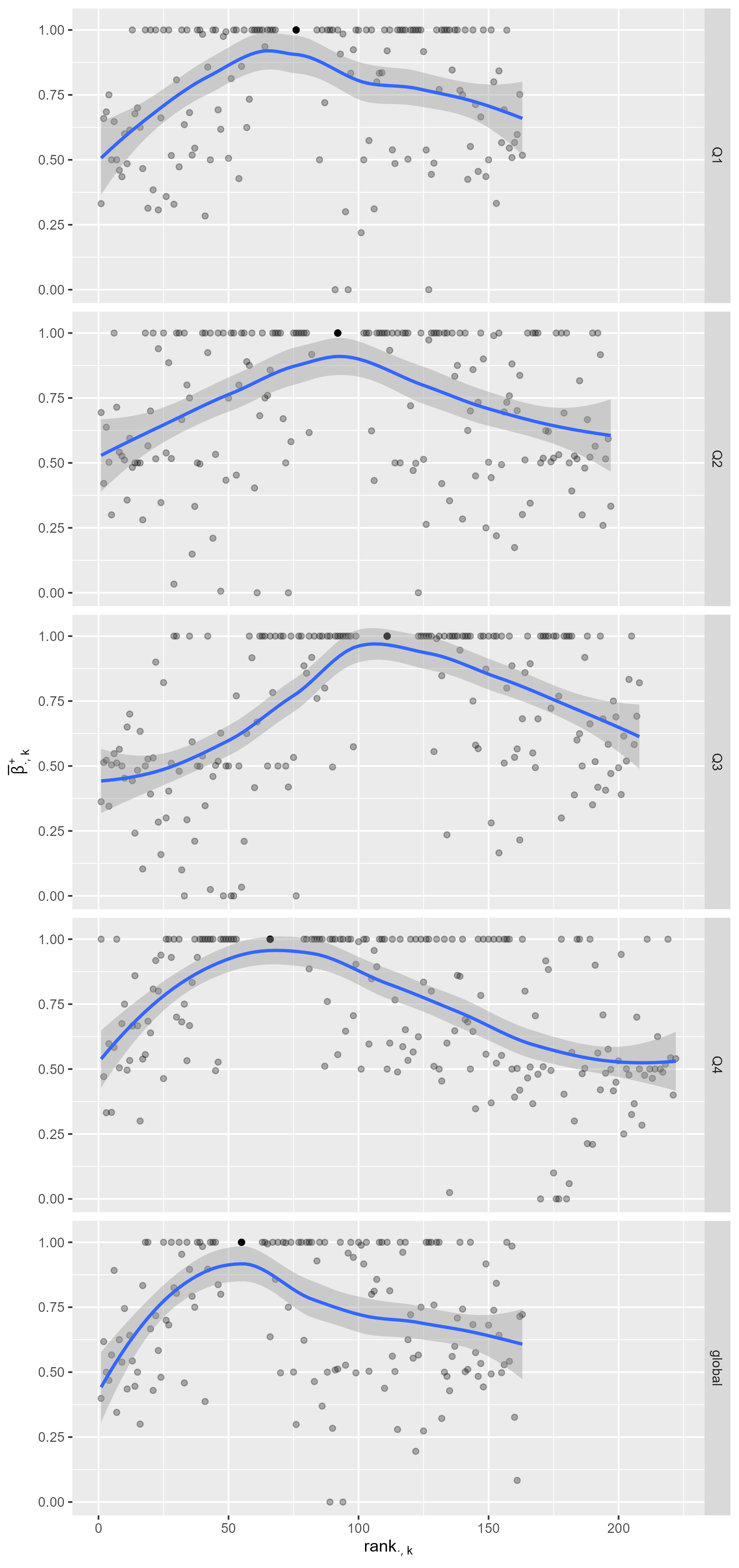}
    \caption{$\bar{\beta}_{\cdot,k}^{+}$ as Function of Attained Rank\\
        The banded lines represent loess estimates of mean with their corresponding standard deviations.
    }
    \label{fig:submission_analysis_rank_long}
\end{figure}

Considering Figures \ref{fig:submission_analysis_long} and \ref{fig:submission_analysis_long_rank_diff} and the simulation results in the theoretical part of this section, one would expect both the best and worst ranking teams to exhibit below-average $\beta^{+}_{m,k}$.
Figure \ref{fig:submission_analysis_rank_long} displays the average $\bar{\beta}^{+}_{m,k}$; $\bar{\beta}^{+}_{\cdot,k}$ in the given quarter and globaly as a function of the attained rank in that period.
Across all quarters and also globally, $\bar{\beta}^{+}_{\cdot,k}$ follows a characteristic inverted ``u'' shape relationship with teams with low $\bar{\beta}^{+}_{\cdot,k}$ being disproportionately represented on either tail of the leaderboard, exactly as predicted.

Table \ref{tab:submission_analysis_median_prob} displays probabilities of attaining rank at least $5$ and $10$ in individual quarters and globally separately for teams with $\bar{\beta}^{+}_{\cdot,k}$ below and above the median $\tilde{\bar{\beta}}^{+}_{\cdot} = q_{0.5}(\bar{\beta}^{+}_{\cdot, :})$ of the respective period.
Teams with below-median $\bar{\beta}^{+}_{\cdot,k}$ are approximately 10 times more likely to secure one of the top 10 positions relative to teams with above-median $\bar{\beta}^{+}_{\cdot,k}$.
This disparity is even more pronounced when focusing on the top 5 leaderboard positions. 
Out of 25 top-5 positions across quarters and globally, only a single one of them is achieved by a team with an above-median $\bar{\beta}^{+}_{\cdot,k}$.

\begin{table}[!htbp]
    \fontsize{6}{6}\selectfont
    \centering
    \begin{filecontents*}{submission_analysis_median_prob.csv}
"period","IR","threshold","p5_FALSE","p5_TRUE","p10_FALSE","p10_TRUE"
"Q1",-0.0535575574803376,0.86,0.0609756097560976,0,0.121951219512195,0
"Q2",-0.626677240331909,0.816161616161616,0.0505050505050505,0,0.0909090909090909,0.0102040816326531
"Q3",-1.90058617821527,0.820672549019608,0.0480769230769231,0,0.0961538461538462,0
"Q4",4.72322757690166,0.8,0.0357142857142857,0.00909090909090909,0.0714285714285714,0.0181818181818182
"global",0.453469805791191,0.8,0.0602409638554217,0,0.108433734939759,0.0125
\end{filecontents*}
\pgfplotstabletypeset[
    col sep = comma,
    ignore chars={"},
    every head row/.style={before row={%
    \toprule
    \multicolumn{13}{c}{\hspace{85pt} $ P(rank_{\cdot,k}\leq 5)  $ \hspace{30pt} $ P(rank_{\cdot,k}\leq 10)  $}\\
    },after row=\midrule},
    every last row/.style={after row=\bottomrule},
    display columns/0/.style={string type, , column type/.add={}{@{\hspace{1em}}},column name={period}},
    display columns/1/.style={dec sep align, precision = 2, fixed, fixed zerofill=true, column type/.add={}{@{\hspace{1em}}}, column name={$IR_{M6\_dummy}$}},
    display columns/2/.style={dec sep align, precision = 3, fixed, fixed zerofill=true, column name={$ \tilde{\bar{\beta}}^{+}_{\cdot}$}},
    display columns/3/.style={dec sep align, precision = 3, fixed, fixed zerofill=true, column name={$ \bar{\beta}^{+}_{\cdot,k}<\tilde{\bar{\beta}}^{+}_{\cdot}$}},
    display columns/4/.style={dec sep align, precision = 3, fixed, fixed zerofill=true, column name={$ \bar{\beta}^{+}_{\cdot,k}\geq\tilde{\bar{\beta}}^{+}_{\cdot}$}},
    display columns/5/.style={dec sep align, precision = 3, fixed, fixed zerofill=true, column name={$ \bar{\beta}^{+}_{\cdot,k}<\tilde{\bar{\beta}}^{+}_{\cdot}$}},
    display columns/6/.style={dec sep align, precision = 3, fixed, fixed zerofill=true, column name={$ \bar{\beta}^{+}_{\cdot,k}\geq\tilde{\bar{\beta}}^{+}_{\cdot}$}}
    ]{submission_analysis_median_prob.csv}
    \caption{$P(rank_{\cdot,k}\leq q)$ of Below/Above Median $\bar{\beta}_{\cdot,k}^{+}$\\
        Probabilities of attaining top rank for teams with mean $\bar{\beta}_{\cdot,k}^{+}$ below/above the median $\tilde{\bar{\beta}}^{+}_{\cdot}$ of the respective period.
    }
    \label{tab:submission_analysis_median_prob}
\end{table}

This is indicative that strategic considerations might have indeed influenced ranking on either tail of the leaderboard, similarly as in the controlled environment of our simulations.
Importantly, these results are not driven by asset returns being negative on average in the respective periods.
As demonstrated by column $IR_{M6\_dummy}$ of Table \ref{tab:submission_analysis_median_prob} which shows the performance of an equal weighted long portfolio, in terms of ranking, it seems to be advantageous to opt for $\bar{\beta}^{+}_{\cdot,k}<\tilde{\bar{\beta}}^{+}_{\cdot}$ irrespective of whether the returns in the given period are positive or negative.

\section{Discussion \& Conclusions}\label{section:conclusions}

To assess the extent to which rankings can be attributed to luck, we employed the WYY test of \citet{wrightTestEqualityMultiple2014} of equality of expected Sharpe ratios across multiple portfolios. 
While the canonical WYY test with analytical critical values tends to over-reject when applied to a large number of teams over the span of a single year, these finite sample distortions can be mitigated by replacing the critical values with bootstrap counterparts.
Applying the WYY test with bootstrap critical values to the portfolio weights submitted by teams yields a p-value of $0.926$, indicating that Sharpe ratios as extreme as those observed in the competition are still compatible with the null hypothesis that none of the teams are capable of consistently outperforming the benchmark, thus not indicative of a violation of the EMH.

This conclusion is corroborated by analyzing directly the stylized model of the competition.
Even when teams submit their portfolios completely at random, one observes equally extreme tail behavior, with the top 1\% of teams achieving an average Sharpe ratio of $29.82$ with a standard deviation of $3.32$, which indeed encompasses the actual observed 99th quantile of Sharpe ratios, $26.22$.

In the second part of the article, we focus on the strategic aspects of the investment challenge arising from the nonlinear reward structure and the adversarial nature of the challenge. 
To assess the benefits of properly accounting for these aspects, we formulate the problem of optimizing directly for the probability of securing a desired rank (rather than maximizing expected Sharpe ratio) as a dynamic programming problem and numerically solve it using the stylized model of the competition.
The optimal rank-optimization portfolio strategy relies heavily on short positions in order to differentiate one's submissions from the predominantly long positions of other teams, and does so more aggressively when one ranks poorly in order to attempt to recover from otherwise hopeless positions.

By employing such a strategy, a team can measurably improve chances of securing a good rank.
Interestingly, these improved probabilities come at the cost of poor expected Sharpe ratios and a disproportionally high probability of placing at the very bottom of the leaderboard, demonstrating that the task of maximizing the expected Sharpe ratio is not necessarily identical to the task of attaining the top rank in the investment challenge.
In the stylized model, the probability of securing the top rank when optimizing directly for it is comparable to that of a team who consistently achieves almost double the market returns. 
To ensure robustness, we also allow the proposed rank optimization portfolio to compete in a bootstrap environment where asset returns and submissions of teams were resampled from those submitted in the M6. 
Here, the success rate is even higher, indicating that the good performance is not an artifact of the simplifying assumptions made to construct the stylized model.

The analysis of portfolio weights submitted by teams participating in the M6 aligns with the stylized facts observed in our simulations and bootstrap exercises. 
A high proportion of short positions is associated with a disproportionately high probability of securing both extremely good and extremely poor ranks. 
This observation is even more pronounced when focusing solely on the top-performing teams.
Teams with below-median proportions of long positions are approximately 10 times as likely to secure one of the top 10 ranks in quarterly rankings and the global ranking. 
Out of 25 teams who secured top 5 ranks in either quarterly or global rankings, only one of them did so with an above-median proportion of long positions.

Clearly, this analysis merely demonstrates that teams could have benefited from recognizing the adversarial nature of the competition, not necessarily that they deliberately did so.
However, descriptions of strategies provided by top-ranking teams suggest that many were indeed aware of the strategic aspects of the competition and incorporated them into their decision-making, often in ways more complex than the inevitably simplified variant analyzed in this article. 
Miguel Perez Michaus, who ranked close second in the duathlon challenge and likely missed the first place by a stroke of luck, states:
\begin{quote}
    I mainly used risk expansion/contraction to maximize my chances for the duathlon prize given a solid forecasting score.
    Depending on market conditions and leaderboard projections I made discretionary use of risk neutral positioning (short SHY / long IEF) marginally benefiting from curve inversion while near-zero volatile, long only positioning, index shorting and volatility based stock selection,...\citep{michausFinQBoostMachineLearning2023}
\end{quote}
Similarly, Shrish, who ranked first in the 4th quarter of the forecasting challenge, emphasized the goal of optimizing decisions to attain a certain rank rather than absolute performance, albeit in the context of the forecasting challenge:
\begin{quote}
    I submitted a baseline forecast for the last month of the quarter to limit the RPS score to 0.16. 
    As a result, my average RPS score was 0.15370, which was satisfactory based on historical data from previous quarters, and gave me a good chance of winning. \citep{shrishWinning6000Few2023}
\end{quote}
Such behavior among teams is not necessarily motivated by mercenary motives of maximizing expected monetary payoff, but can equally be driven simply by the intrinsic desire to win. 
This is demonstrated by \citet{dijkRankMattersImpact2014}, who show that such behavior arises to the same degree even in the absence of monetary rewards, purely because of social competition. 
Consequently, even more equitably distributed prizes would likely not mitigate these incentives.

It must be stressed that this analysis, by its very construction, cannot make claims about how individual teams achieved their results or downplay the excellent Sharpe ratios they attained.
Nor can it rule out with certainty the possibility that some teams might indeed be capable of consistently outperforming the benchmark.
It merely shows that, taking the results of the investment challenge as a whole, there does not appear to be sufficient evidence to suggest that the extreme Sharpe ratios observed are beyond what one would expect given the large number of teams competing in M6.
Thus, they should not be interpreted as evidence against the EMH.
We recommend exercising caution in placing undue focus solely on solutions from top-performing teams as those, by definition, might be more fortunate than others and might be more likely to utilize strategies which lead to increased probabilities of attaining extreme ranks.
Instead, a more holistic approach focusing on how different factors affect performance across the entire leaderboard might be preferred.

\bibliographystyle{elsarticle-harv}\biboptions{authoryear}
\bibliography{Library.bib}

\appendix

\section{Estimation}

\subsection{Returns Distribution}\label{appendix:returns_distribution}

We estimate $\widehat{\sigma}_{r,r}$ and $\widehat{\sigma}_{r,r'}$ by MLE.
An explicit estimator for the covariance matrix under compound symmetry \citep[see][p. 95]{seberMultivariateObservations1984} is
\begin{equation}
    \widehat{\sigma}_{r,r}=\dfrac{1}{I}\sum_{i=1}^{I}s_{i,i},
\end{equation}
\begin{equation}
    \widehat{\sigma}_{r,r'}=\dfrac{1}{I*(I-1)}\sum_{i\neq i'}^{R}s_{i,i'},
\end{equation}
where $s_{i,i'}$ is the conventional unbiased covariance estimate.

\subsection{Baseline Portfolio}\label{appendix:baseline_portfolio}

We estimate $\theta=\{ n_{+1}, n_{0}, n_{-1}\}$ using the method of simulated moments \citep{mcfaddenMethodSimulatedMoments1989}.
The estimation is performed by matching the mean and the kurtosis of $IR_{T_{m},:}$ for any given submission $m$ with those observed using weights simulated via Eq. \ref{eq:baseline_portfolio}.
We opt for kurtosis since the variance is non-informative given that $IR_{T_{m},k}$ is already standardized.
Let us denote the data from from interval $m$ as $\mathcal{D}_{m}=\{w_{:,m,:}, \{r_{:,t}\}_{t\in T_{m}}\}$.
The moment function is defined as
\begin{equation}
    g(\mathcal{D}_{m},\theta)=
    \begin{pmatrix}
        \dfrac{\left(g_{1}(\mathcal{D}_{m})- \hat{\mu}_{g_{1}}(\theta)\right)^2 - \hat{\sigma}^{2}_{g_{1}}(\theta)}{\sqrt{\hat{\sigma}^{2}_{g_{1}}(\theta)}} \\
        \dfrac{\left(g_{2}(\mathcal{D}_{m})- \hat{\mu}_{g_{2}}(\theta)\right)^2 - \hat{\sigma}^{2}_{g_{2}}(\theta)}{\sqrt{\hat{\sigma}^{2}_{g_{2}}(\theta)}} \\
    \end{pmatrix}
\end{equation}
with
\begin{equation}
    g_{1}(\mathcal{D}_{m}) = K^{-1}\sum_{k=1}^{K}IR_{T_{m},k}
\end{equation}
\begin{equation}
    g_{2}(\mathcal{D}_{m}) = \frac{K^{-1}\sum_{k=1}^{K}(IR_{T_{m},k}-K^{-1}\sum_{k'=1}^{K}IR_{T_{m},k'})^{4}}{\left(K^{-1}\sum_{k=1}^{K}(IR_{T_{m},k}-K^{-1}\sum_{k'=1}^{K}IR_{T_{m},k'})^{2}\right)^{2}}
\end{equation}
and with
\begin{equation}
    \hat{\mu}_{g_{i}}(\theta)=\underset{\tilde{w}_{:,m,:}|\theta}{\widehat{\mathbb{E}}}[g_{i}(\{\tilde{w}_{:,m,:}, \{r_{:,t}\}_{t\in T_{m}}\})]
\end{equation}
\begin{equation}
    \hat{\sigma}_{g_{i}}^{2}(\theta)=\underset{\tilde{w}_{:,m,:}|\theta}{\widehat{\mathbb{V}}}[g_{i}(\{\tilde{w}_{:,m,:}, \{r_{:,t}\}_{t\in T_{m}}\})]
\end{equation}
estimated by repeatedly drawing $\tilde{w}_{:,m,:}$ under $\theta$ via Eq. \ref{eq:baseline_portfolio}.
The estimation itself is performed by minimizing
\begin{equation}
    \left(\dfrac{1}{12}\sum_{m=1}^{12}g(\mathcal{D}_{m},\theta)\right)' W \left(\dfrac{1}{12}\sum_{m=1}^{12}g(\mathcal{D}_{m},\theta)\right)
\end{equation}
through an exhaustive search over $\{n_{+1}, n_{0}, n_{-1}|n_{+1}+n_{0}+n_{-1}=100, n_{0} \neq 100\}$.
The weighting matrix is assumed to be an identity matrix because the moments are already standardized.

\section{Auxiliary Results}\label{appendix:auxiliary_results}

\begin{figure}[!htbp]
    \centering
    \includegraphics[width=1\linewidth]{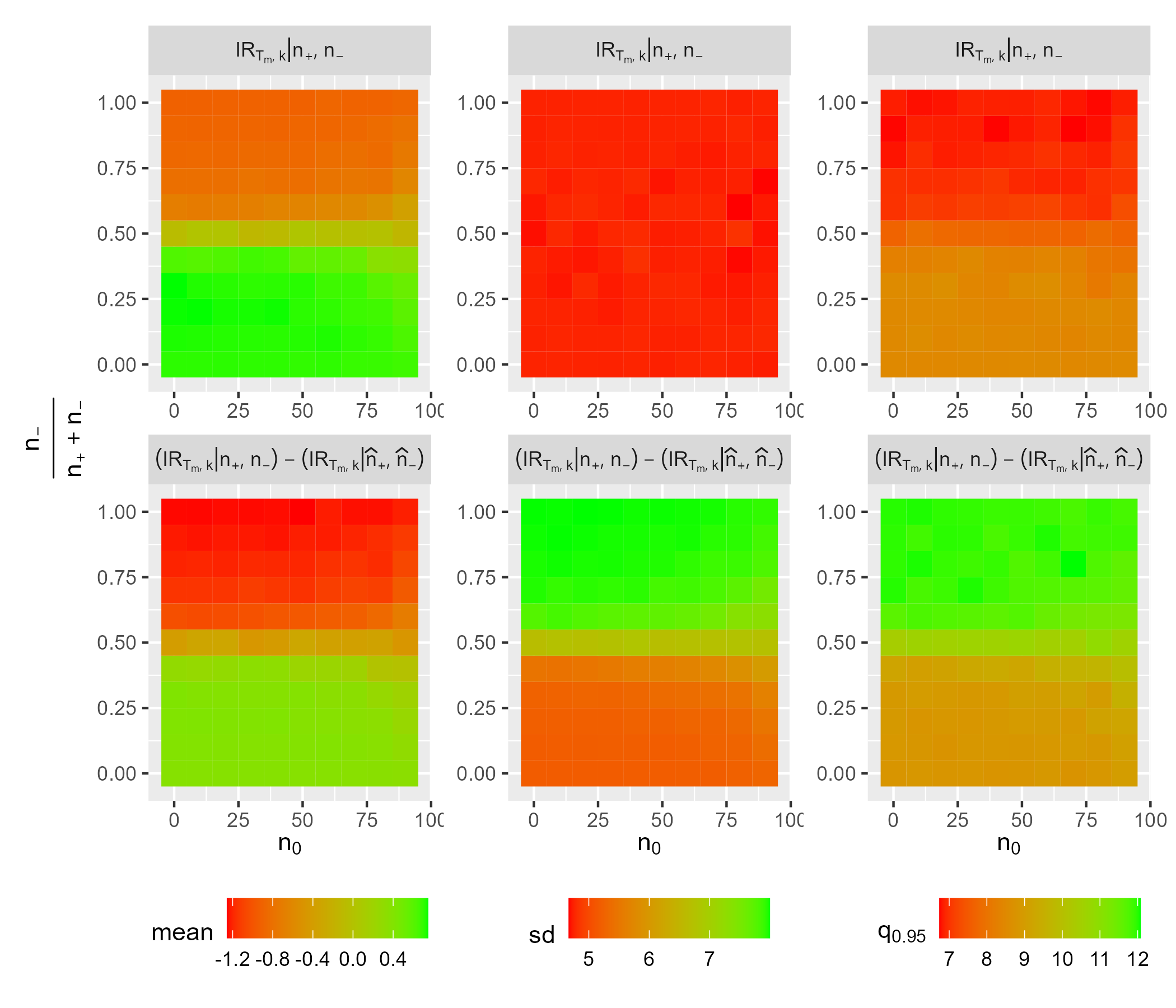}
    \caption{Effects of $n_{+}$, $n_{-}$, and $n_{0}$ on Distribution of $IR_{T_{m},k}$\\
    \footnotesize
    The first row displays the mean, standard deviation, and $95$-th quantile of $IR_{T_{m},k}$ as a function of the number of zero positions $n_{0}$ (horizontal axis) and the proportion of short positions $n_{-}/(n_{+}+n_{-})$ (vertical axis).
    The second row displays analogous plots but for the difference $IR_{T_{m},k}$ and the $IR_{T_{m},k'}$ of the benchmark portfolio under the estimated $\hat{n}_{+}$ and $\hat{n}_{-}$.
    The ratio of negative positions has an adverse effect on the performance of $IR_{T_{m},k}$ in isolation.
    However, when measured relative to the $IR_{T_{m},k'}$ of the benchmark portfolio, taking more short positions increases the standard deviation and upper quantiles.
    The number of zero positions $n_{0}$ seems to have no effect on either $IR_{T_{m},k}$ or their difference because of the standardization.
    }
    \label{fig:position_effects}
\end{figure}

\begin{figure}[!htbp]
    \centering
    \includegraphics[width=1\linewidth]{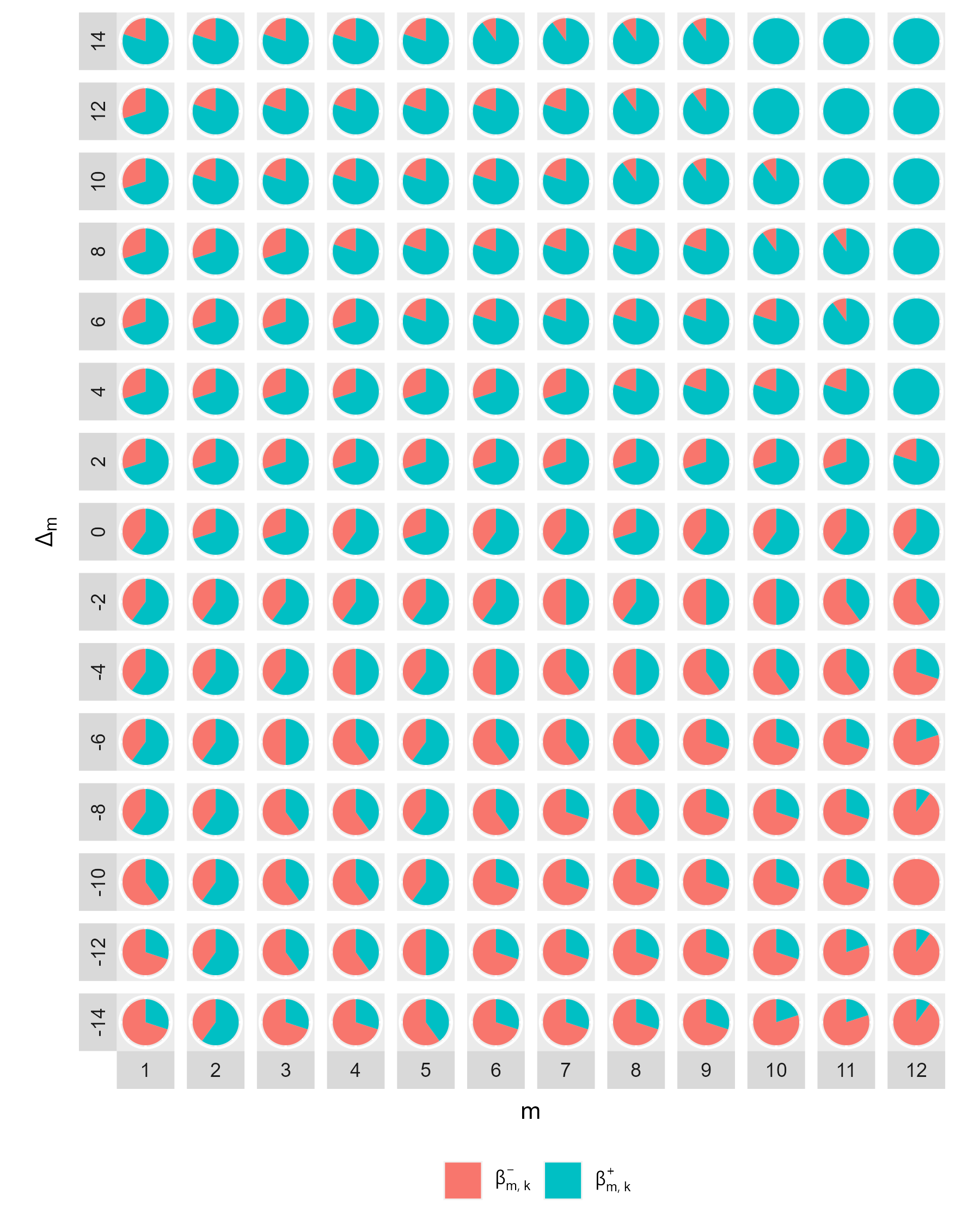}
    \caption{Rank Optimization Portfolio ($q=20$)\\
        \footnotesize
    }
    \label{fig:strategy_diagram_q_20}
\end{figure}

\begin{figure}[!htbp]
    \centering
    \includegraphics[width=0.99\linewidth]{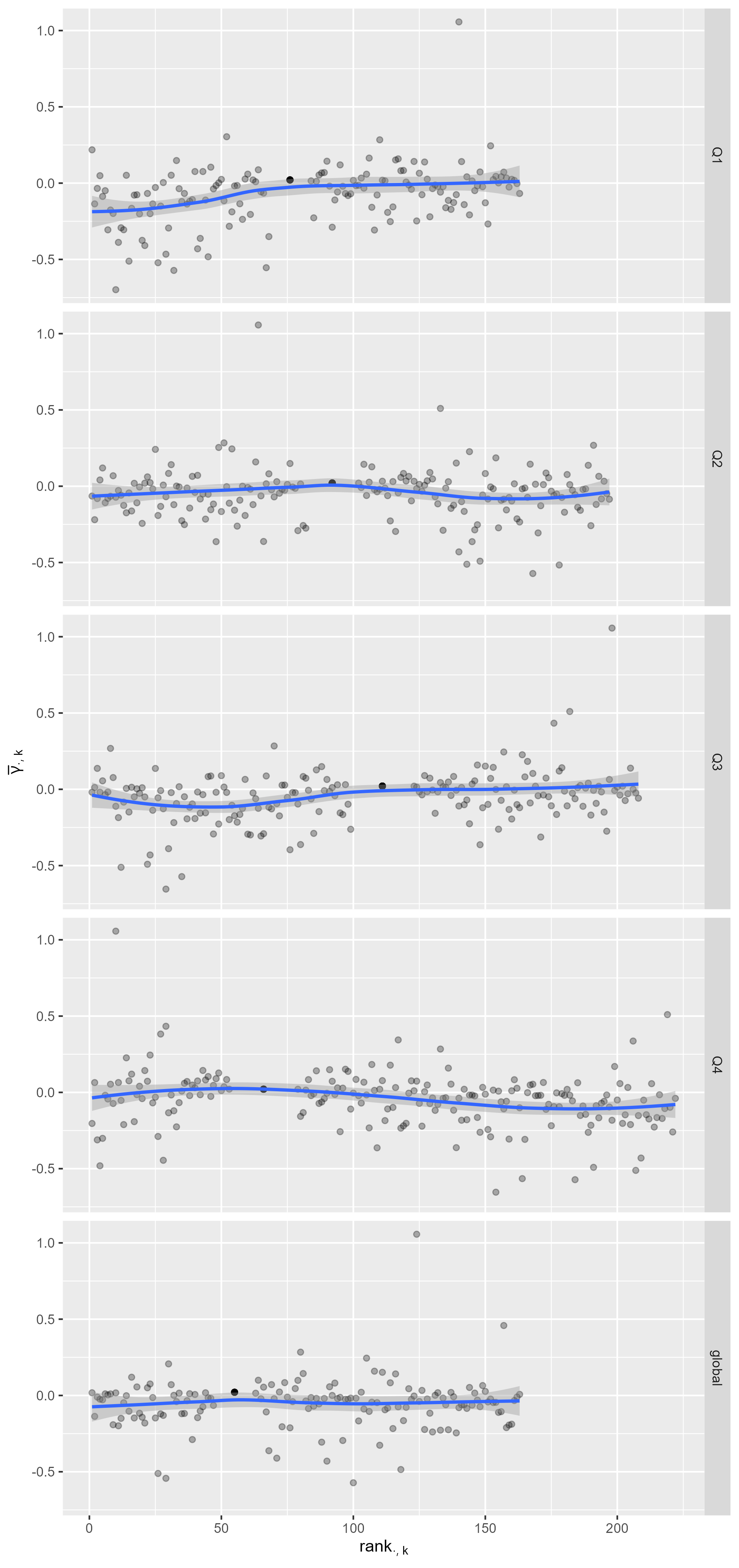}
    \caption{Average $\gamma_{m,k}$; $\bar{\gamma}_{\cdot,k}$ as Function of Attained Rank\\
        Parameter $\gamma_{m,k}=\sum_{i = 1}^{I} \frac{w_{i,m,k}}{\sum_{i = 1}^{I}|w_{i,m,k}|}\widehat{\gamma}_{i}$ with $\widehat{\gamma}_{i}$ being the sample skewness computed over the duration of the competition  measures the exposure of team $k$ to positively (resp. negatively) skewed assets.
        The banded lines represents loess estimates of mean with its corresponding standard deviations.
    }
    \label{fig:submission_analysis_rank_skew}
\end{figure}

\end{document}